\documentclass[
  aps,
  reprint,
  prb,
  showpacs,
  superscriptaddress,
  longbibliography,
]{revtex4-1}
\pdfoutput=1

\usepackage{graphicx}
\usepackage{amsmath}
\usepackage{amssymb}
\usepackage{amsfonts}
\usepackage{color}
\usepackage{hyperref}

\newcommand{\vertexop}[1]{\mathcal{V}^{\alpha}_{s_{#1}}(z_{#1})}
\newcommand{\phifull}[1]{\phi(z_{#1}, \bar{z}_{#1})}
\newcommand{\spherepos}[1]{\mathbf{n}_{\Omega_{#1}}}
\newcommand{\Ci}{\mathrm{Ci}}
\newcommand{\Si}{\mathrm{Si}}
\newcommand{\zz}[2]{\langle \sigma^z_{#1} \sigma^z_{#2} \rangle}
\newcommand{\Sfb}{\frac{1}{4 \pi} \int dz d\bar{z} \partial_z \phi(z, \bar{z}) \partial_{\bar{z}} \phi(z, \bar{z})}
\newcommand{\Galpha}{G^{\alpha}}
\newcommand{\Salpha}{S_{\alpha}}
\newcommand{\Zalpha}{\mathcal{Z}_{\alpha}}
\newcommand{\evec}[1]{\mathbf{e}_{#1}}
\newcommand{\gaux}[1]{\cos\left(#1\right) \Ci\left(#1\right) + \sin\left(#1\right) \left(\Si\left(#1\right) - \frac{\pi}{2}\right)}
\newcommand{\Mmat}[2]{\delta_{#1, #2} - 2 \alpha \ln\left(d_{#1, #2} + \delta_{#1, #2}\right)}
\newcommand{\Trmat}[2]{\delta_{#1, #2} - \delta_{#1, r}}
\newcommand{\Gammamat}{T_r \left(T_r^t M T_r + \evec{r} \evec{r}^t\right)^{-1} T_r^t - \mathbb{I}}
\newcommand{\aldef}{\frac{1}{2} (\zz{l+2}{1} + \zz{l+1}{1})}
\newcommand{\ptitle}{Effective description of correlations for states obtained from conformal field theory}
\newcommand{\kwds}{strongly correlated systems; fractional quantum Hall effect; conformal field theory}
\newcommand{\MPQ}{Max-Planck-Institut f{\"u}r Quantenoptik, Hans-Kopfermann-Stra{\ss}e\ 1, D-85748 Garching, Germany}

\hypersetup{
  pdftitle={\ptitle},
  pdfkeywords={\kwds},
  pdfauthor={Benedikt Herwerth, Germ{\'a}n Sierra, J. Ignacio Cirac, Anne E. B. Nielsen},
  pdfsubject={},
  pdfcreator={pdflatex},
  colorlinks,
  linkcolor=blue,
  citecolor=blue,
  urlcolor=blue
}

\begin{document}
\author{Benedikt Herwerth}
\email{benedikt.herwerth@mpq.mpg.de}
\affiliation{\MPQ}
\author{Germ\'an Sierra}
\affiliation{Instituto de F{\'i}sica Te{\'o}rica, UAM-CSIC, Madrid, Spain}
\author{J. Ignacio Cirac}
\affiliation{\MPQ}
\author{Anne E. B. Nielsen}
\affiliation{Max-Planck-Institut f{\"u}r Physik komplexer Systeme, D-01187 Dresden, Germany}
\affiliation{\MPQ}
\affiliation{Department of Physics and Astronomy, Aarhus University, DK-8000 Aarhus C, Denmark}

\title{\ptitle}

\begin{abstract}
We study states of one- and two-dimensional spin systems that are
constructed as correlators within the conformal field theory of
a massless, free boson.
In one dimension, these are good variational wave functions for XXZ spin
chains
and they are similar to lattice Laughlin states in two dimensions.
We show that their $zz$ correlations are determined
by a modification of the original free-boson theory.
An expansion to quadratic order leads to a solvable,
effective theory for the correlations in these states. Compared to
the massless boson, there is an additional term
in this effective theory that explains the behavior of the correlations:
a polynomial decay in one dimension and at the edge of a two-dimensional system and
an exponential decay in the bulk of a two-dimensional system.
We test the validity of our approximation by comparing it to Monte Carlo
computations.
\end{abstract}


\maketitle

\section{Introduction}
Conformal field theory (CFT) in $1+1$ dimensions is characterized by
the absence of an intrinsic energy scale and has applications in
various fields including string theory, statistical mechanics, and
condensed matter physics. In the latter case, CFT arises as the continuous,
field theoretical description of low-energy degrees of freedom of critical
lattice models in one spatial dimension (1D). Traditionally, this
connection is established by explicitly taking the continuum limit of
a lattice Hamiltonian, potentially integrating out high-energy modes
through renormalization, and mapping operators of the original
problem to continuous fields. 

In the past decades, another approach in applying CFT to problems
in condensed matter physics was established. In this case, CFT is used
to construct variational wave functions for interacting many-body
systems, both in 1D and in two spatial dimensions (2D).
This idea reaches back to Laughlin's description of the
fractional quantum Hall (FQH) effect\cite{Laughlin1983} and the realization
that the Laughlin wave function and its generalizations are
correlators in a CFT\cite{Moore1991}.  Such a construction was then
also established for systems on a lattice\cite{Cirac2010}, where the
corresponding states can be understood as generalizations of matrix
product states to the case of an infinite bond dimension\cite{Cirac2010}.
Within this framework, many interesting states and model systems
were obtained including
1D critical and
FQH lattice states\cite{Nielsen2012, Tu2013, Glasser2015, Glasser2016, Hackenbroich2017}.

The fact that variational wave functions constructed from CFT
provide good descriptions of FQH states may seem surprising:
In the bulk, a FQH system is gapped and has exponentially
decaying correlations whereas a CFT is gapless and CFT correlators
decay polynomially.
This raises the question of how the CFT influences the
properties of corresponding states.

In the past years,
it was realized that the entanglement entropy and
spectra carry signatures of the underlying CFT:
In 1D, the entanglement entropy of a CFT exhibits a
logarithmic violation of the area law
and this characteristic behavior was observed in
corresponding lattice systems\cite{Vidal2003, Calabrese2004}.
Furthermore, the energy levels of CFTs were found
in the entanglement spectra of 1D\cite{Laeuchli2013} and of FQH states
\cite{Li2008, Sterdyniak2012, Dubail2012, Rodriguez2012}.
In addition to its entanglement properties, the CFT nature of
a state also affects its correlations. While both CFT correlators
and those of a critical 1D state decay polynomially, the
relation is more intricate for 2D states, which decay
polynomially only at the edge of a 2D system.
Assuming that their bulk correlations decay exponentially, it was shown
in Ref.~\onlinecite{Dubail2012a} that edge correlations of a FQH state
follow from those of the underlying CFT. So far, our knowledge
concerning bulk correlations relies on numerical computations,
mostly of Monte Carlo type\cite{Caillol1982, Nielsen2012}.

The purpose of this paper is to clarify the relation between
the correlations in a state and the CFT
it is constructed from. We express these correlations
as expectation values in a field theory that is a modification
of the original CFT. This explains the different types of
correlations, in particular those in the bulk of a 2D system
as emerging from a mass-like perturbation of the CFT.

The idea of treating the effect
of a FQH wave function as a perturbation of the underlying CFT was introduced
in a general context in Refs.~\onlinecite{Read2009, Dubail2012a}.
The screening property, i.e. exponentially decaying bulk correlations,
was assumed in Ref.~\onlinecite{Dubail2012a}.
Here, we provide analytical evidence for screening
for a specific type of FQH states: By an expansion
of the exact representation of the correlations,
we derive an effective action that
has a mass-like term in the bulk of the system, thus giving
rise to exponentially decaying bulk correlations.

Concretely, we study a set of states that are constructed from the
CFT of a free, massless boson. These are defined for a system of
spin-$\frac{1}{2}$ degrees of freedom on a lattice. In addition to
the lattice position, they have one parameter corresponding to
the scaling dimension of CFT primary fields.
In 1D, these states are good descriptions
of ground states of the XXZ spin chain in the half-filled sector~\cite{Cirac2010}
and they resemble lattice Laughlin states in 2D~\cite{Nielsen2012}.
For the $zz$ spin correlations,
we derive an exact path integral representation
and then truncate it to quadratic order to obtain an effective, free
theory.
In contrast to the free-boson CFT, the action of this effective theory
has a mass-like term, which corresponds to the scaling dimension of the primary
fields that define our states. Within the approximation of a free field
theory, the correlations on lattices can be computed efficiently for large
system sizes. Taking a continuum limit,
we furthermore obtain analytical results for the approximate $zz$
correlations. Our calculations provide evidence for
a power-law decay in 1D and at the edge of a 2D system and
an exponential decay in the bulk of a 2D system. This behavior
is a consequence of the mass-like term in the effective theory.

We test the validity of our approximation in
1D and 2D by
comparing the results to Monte Carlo computations. Our analysis
shows that the approximation is better for smaller scaling dimensions.
Furthermore, it is applicable
for a larger range of scaling dimensions in the case of
a 2D system compared to a 1D system.

This paper is structured as follows. Sec.~\ref{sec:spin-states-from-cft}
describes the construction of states from the free-boson CFT and
Sec.~\ref{sec:effective-description-and-expansion} derives
an effective description of their $zz$ correlations.
In Sec.~\ref{sec:solution-schemes}, we describe our solution
to the effective theory within a continuum limit and on lattices.
The results of our approximation are compared to the exact correlations
in Sec.~\ref{sec:quality-of-approximation}.

\section{Spin states from conformal fields}
\label{sec:spin-states-from-cft}
We consider a free, massless bosonic field $\phi(z, \bar{z})$ in one spatial dimension
with the Euclidean action
\begin{align}
\label{eq:action-free-massless-boson}
S_0[\phi] &= \Sfb,
\end{align}
where $z$ and $\bar{z}$ are coordinates in the complex plane. The field $\phi(z, \bar{z})$ decomposes
into a chiral part $\phi(z)$ and an anti-chiral part $\bar{\phi}(\bar{z})$
according to $\phi(z, \bar{z}) = \phi(z) + \bar{\phi}(\bar{z})$.
The chiral and anti-chiral sectors of the free-boson theory define
CFTs with central charges $c=1$ and $\bar{c}=1$,
respectively. The vertex operators
\begin{align}
:e^{i \sqrt{\alpha} s \phi(z)}:
\end{align}
are chiral primary fields with a scaling dimension of $\alpha/2$.\cite{DiFrancesco1997}
Here, $\alpha$ is a positive real parameter, $s \in \{-1, 1\}$, and the
colons denote normal ordering.
We consider $N$ vertex operators at distinct positions $z_j$ in the complex plane:
\begin{align}
\vertexop{j} = \chi_{j, s_j} :e^{i \sqrt{\alpha} s_j \phi(z_j)}:,
\end{align}
where $\chi_{j, s_j}$ are $2 N$ phase factors ($|\chi_{j, s_j}|=1$)
that we leave unspecified
since the correlations computed in this work are independent of
the choice of $\chi_{j, s_j}$.
The CFT correlator of these operators defines the spin states $\psi_{\alpha}$ through
\begin{align}
| \psi_{\alpha} \rangle &= \sum_{s_1, \dots, s_N} \psi_{\alpha}(s_1, \dots, s_N) | s_1, \dots, s_N \rangle,\\
\label{eq:definition-of-wave-function}
\psi_{\alpha}(s_1, \dots, s_N) &= \langle \vertexop{1} \dots \vertexop{N} \rangle.
\end{align}
Here, $s_j \in \{-1, 1\}$,
$|s_1, \dots, s_N\rangle$ is the tensor product of eigenstates
of the spin-$z$ operator $t^z_j$ at position $j$ in the spin-$1/2$
representation $\left(t^z_j | s_j \rangle = \frac{s_j}{2} | s_j \rangle\right)$,
and $\langle \dots \rangle$ denotes the radially ordered vacuum expectation
value in the free-boson CFT.
Evaluating the CFT correlator leads to~\cite{DiFrancesco1997}
\begin{align}
\label{eq:wave-function-psi-alpha}
\psi_{\alpha}(s_1, \dots, s_N) = \delta_{\mathbf{s}} \prod_{j=1}^N \chi_{j, s_j} \prod_{i < j}^N (z_i - z_j)^{\alpha s_i s_j},
\end{align}
where $\delta_{\mathbf{s}} = 1$ if $\sum_{j=1}^N s_j = 0$ and $\delta_{\mathbf{s}} = 0$ otherwise.
Due to the charge neutrality condition $\sum_{j=1}^N s_j = 0$,
the states $\psi_{\alpha}$ have a vanishing $z$ component of the total spin.

For a 2D system, $\psi_{\alpha}$ is similar to the Laughlin lattice state with
$\nu = \frac{1}{4 \alpha}$ particles per flux\cite{Nielsen2012}.
In particular, $\alpha=\frac{1}{4}$ corresponds to an integer quantum Hall state
with one particle per flux
and $\alpha=\frac{1}{2}$ to the Kalmeyer-Laughlin FQH state with $1/2$ particle
per flux. When $4 \alpha$ is not an integer, the states $\psi_{\alpha}$
can be thought of as generalizations of FQH lattice states.

\section{Effective description of correlations}
\label{sec:effective-description-and-expansion}
In this section, we derive an effective description of the $zz$
correlations in the states $\psi_{\alpha}$ in terms of a free field
theory. The $zz$ correlations between lattice points $i$
and $j$ are defined as
\begin{align}
  \label{eq:definition-of-zz-correlations}
  \zz{i}{j} &=
  4 \frac{\langle \psi_{\alpha} | t^z_i t^z_j | \psi_{\alpha} \rangle}
                  {\langle \psi_{\alpha} | \psi_{\alpha} \rangle}\\
&  = \frac{\sum_{s_1, \dots, s_N} s_i s_j |\psi_{\alpha}(s_1, \dots, s_N)|^2}
  {\sum_{s_1, \dots, s_N} |\psi_{\alpha}(s_1, \dots, s_N)|^2}.
\end{align}
In the following, it is assumed that the
sites $i$ and $j$ are distinct since
the correlator for $i = j$ can be evaluated trivially, $\zz{i}{i} = 1$.
We first derive an exact representation of $\zz{i}{j}$,
which we then truncate to an effective, quadratic theory.

\subsection{Exact field theory representation of correlations}
Let us first consider the normalization
$\langle \psi_{\alpha} | \psi_{\alpha} \rangle$ in Eq.~\eqref{eq:definition-of-zz-correlations}.
Using the form of the wave function of Eq.~\eqref{eq:wave-function-psi-alpha},
$\zz{i}{j}$ can be written
as a vacuum expectation value of vertex operators in the complete free-boson theory (chiral
and anti-chiral):
\begin{align}
&\langle \psi_{\alpha} | \psi_{\alpha} \rangle\notag \\
&\quad= \sum_{s_1, \dots, s_N} |\psi_{\alpha}(s_1, \dots, s_N)|^2 \\
&\quad= \sum_{s_1, \dots, s_N} \delta_{\mathbf{s}} \prod_{i < j} |z_i - z_j|^{2 \alpha s_i s_j} \\
&\quad= \sum_{s_1, \dots, s_N} \langle :e^{i \sqrt{\alpha} s_1 \phifull{1}}: \dots  :e^{i \sqrt{\alpha} s_N \phifull{N}}: \rangle,
\label{eq:normsquared-as-CFT-correlator}
\end{align}
where $\delta_{\mathbf{s}} \prod_{i < j} |z_i - z_j|^{2 \alpha s_i s_j}$
was written as the correlation function of $N$ vertex operators
$:e^{i \sqrt{\alpha} s_j \phifull{j}}:$.
Note that the condition imposed by $\delta_{\mathbf{s}}$
is implicitly contained
in Eq.~\eqref{eq:normsquared-as-CFT-correlator} since the
correlator of vertex operators vanishes unless $s_1 + \dots + s_N = 0$.

Carrying out each of the sums over $s_j$,
\begin{align}
\sum_{s_j \in \{-1, 1\}} :e^{i \sqrt{\alpha} s_j \phifull{j}}: = 2 :\cos\left(\sqrt{\alpha} \phifull{j} \right):,
\end{align}
we obtain
\begin{align}
\langle \psi_{\alpha} | \psi_{\alpha} \rangle &=
2^N \langle \prod_{k=1}^N :\cos\left(\sqrt{\alpha} \phifull{k} \right): \rangle.
\end{align}
For the numerator in Eq.~\eqref{eq:definition-of-zz-correlations}, we additionally use
\begin{align}
\sum_{s_j \in \{-1, 1\}} s_j :e^{i\sqrt{\alpha} s_j \phifull{j}}: &=  2 i :\sin\left(\sqrt{\alpha} \phifull{j}\right):
\end{align}
and obtain
\begin{align}
&4 \langle \psi_{\alpha} | t^z_i t^z_j | \psi_{\alpha} \rangle \notag \\
&= \sum_{s_1, \dots, s_N} s_i s_j \langle :e^{i \sqrt{\alpha} s_1 \phifull{1}}: \dots  :e^{i \sqrt{\alpha} s_N \phifull{N}}: \rangle\\
&= -2^N \langle :\sin\left(\sqrt{\alpha} \phifull{i}\right): :\sin\left(\sqrt{\alpha} \phifull{j}\right):  \times \notag \\
&\quad\phantom{-2^N} \times \prod_{k (\neq i, j)}^N :\cos\left(\sqrt{\alpha} \phifull{k}\right): \rangle,
\end{align}
where $k (\neq i, j)$ denotes all indices $k$ that are distinct from
$i$ and $j$.
Therefore, the expression for the $zz$ correlations becomes
\begin{widetext}
\begin{align}
\label{eq:zz-correlations-exact-expression-before-path-integral}
\zz{i}{j} &= -\frac{\langle :\sin\left(\sqrt{\alpha} \phifull{i}\right): :\sin\left(\sqrt{\alpha} \phifull{j}\right):
\prod_{k (\neq i, j)}^N :\cos\left(\sqrt{\alpha} \phifull{k}\right):\rangle}
{\langle \prod_{k=1}^N :\cos\left(\sqrt{\alpha} \phifull{k}\right):\rangle \rangle}.
\end{align}
As we show in Appendix~ \ref{sec:appendix-normal-ordering}, one can drop the normal ordering
in this expression since this changes the numerator and denominator by the same constant factor.
Thus, the path integral representation of
Eq.~\eqref{eq:zz-correlations-exact-expression-before-path-integral} is given by
\begin{align}
  \label{eq:zz-correlations-exact-expression-path-integral}
  \zz{i}{j} &= -\frac{\int \mathcal{D} \phi \tan\left(\sqrt{\alpha} \phifull{i}\right) \tan\left(\sqrt{\alpha} \phifull{j}\right) \cos\left(\sqrt{\alpha} \phifull{1} \right) \dots \cos\left(\sqrt{\alpha} \phifull{N}\right) e^{-S_0[\phi]}}
  {\int \mathcal{D} \phi \cos\left(\sqrt{\alpha} \phifull{1} \right) \dots \cos\left(\sqrt{\alpha} \phifull{N}\right) e^{-S_0[\phi]}}.
\end{align}
\end{widetext}
This expression determines the exact $zz$ correlations in $\psi_{\alpha}$.

\subsection{Effective theory for correlations}
\label{sec:effective-theory}
The starting point of our approximation is to expand
the path integral representation of
Eq.~\eqref{eq:zz-correlations-exact-expression-path-integral} to quadratic
order in $\phi(z, \bar{z})$ around $0$. This is motivated by the following
observation: For large $N$,  the contribution of the integrand in
Eq.~\eqref{eq:zz-correlations-exact-expression-path-integral}
is only significant for field configurations that have
$\cos(\sqrt{\alpha} \phifull{j}) \approx \pm 1$ at all positions $z_j$.
At the same time, the massless, free-boson action $S_0[\phi]$ suppresses field
configurations that change rapidly through the derivative term.
Therefore, the fields dominating the path integral are those for which
$\sqrt{\alpha} \phi(z, \bar{z})$ is near the same extremum of cosine
for all positions $z$.  Since $S_0[\phi]$ is invariant under a constant
shift of the field value,
$\phi(z, \bar{z}) \to \phi(z, \bar{z}) + \text{const.}$,
we can focus on the case $\phi(z, \bar{z}) \approx 0$.
The expansion around the extremum of the cosine function is analogous
to Kosterlitz and Thouless's treatment of
the XY model~\cite{Kosterlitz1973, Kosterlitz1974}. We are, however,
not taking into account terms that would correspond to vortex configurations
in the XY model.

Hence, we expand
$\cos(\sqrt{\alpha} \phi(z, \bar{z})) \sim e^{-\frac{\alpha}{2}
\phi(z, \bar{z})^2}$
and
$\tan(\sqrt{\alpha} \phi(z, \bar{z})) \sim \sqrt{\alpha}
\phi(z, \bar{z})$
in Eq.~\eqref{eq:zz-correlations-exact-expression-path-integral} and
obtain
\begin{align}
\label{eq:zz-correlations-approximate-expression-path-integral}
  \zz{i}{j} &\approx - \alpha \frac{\int \mathcal{D} \phi \; \phifull{i} \phifull{j} e^{-S_{\alpha}[\phi]}}
  {\int \mathcal{D} \phi  e^{-S_{\alpha}[\phi]}},
\end{align}
where
\begin{align}
\label{eq:quadratic-action}
S_{\alpha}[\phi] &= \Sfb + \frac{\alpha}{2} \sum_{j=1}^N \phifull{j}^2.
\end{align}
The quadratic action $S_{\alpha}$ provides an effective theory that approximately
describes the $zz$ correlations in the state $\psi_\alpha$.
Compared to the action of the free, massless
boson, it has an additional mass-like term at the positions of the lattice.

\section{Solution schemes for 1D and 2D lattices}
\label{sec:solution-schemes}
In the following, we describe two solution schemes to the
quadratic action $S_\alpha$ of Eq.~\eqref{eq:quadratic-action}, which determines the
approximate $zz$ correlations in the state $\psi_{\alpha}$. 
The first scheme consists of taking a continuum limit of
the lattice. This further simplification allows us
to derive analytical results for the approximate $zz$ correlations.
The second scheme keeps the structure of the lattice and is solved numerically.

\subsection{Continuum approximation}
\begin{figure*}[htb]
\centering
\includegraphics[width=1.\linewidth]{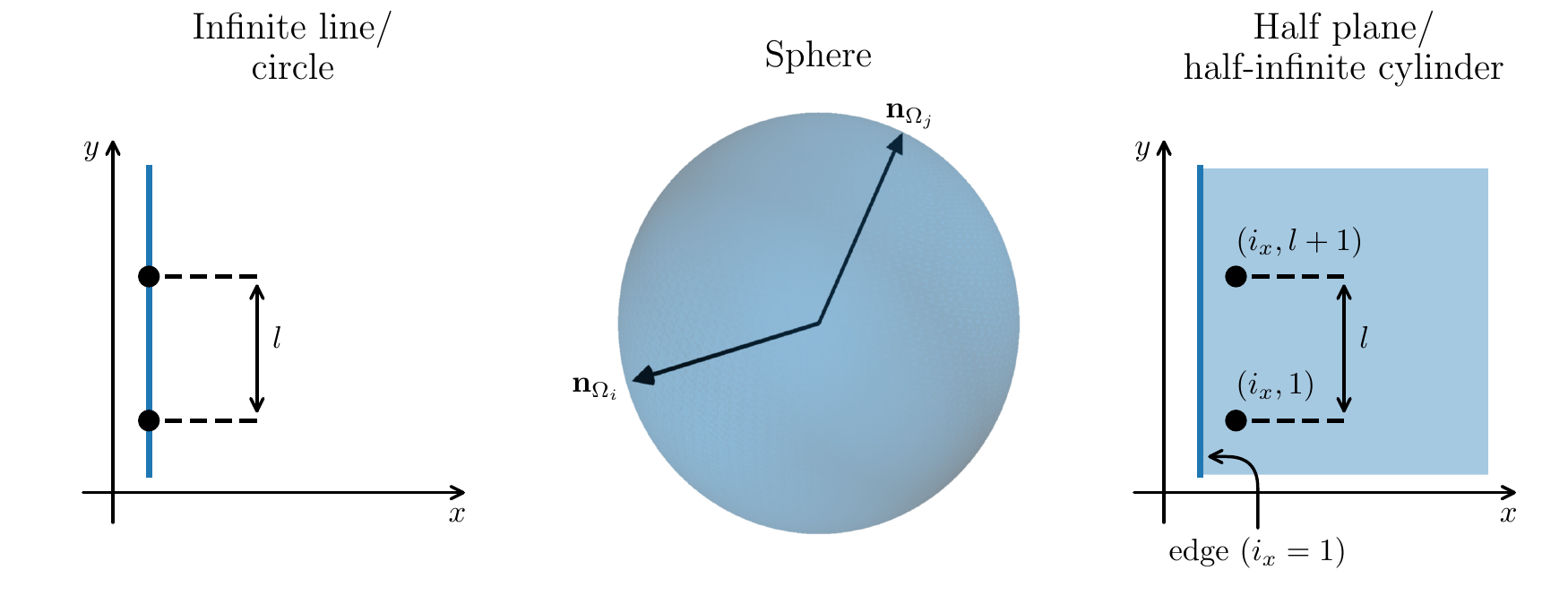}
\caption{(Color online) Systems for the computation of the $zz$ correlations
in the continuum approximation. The spins are located in the blue regions.
In the case of the circle and the half-infinite cylinder, periodic boundary
conditions are imposed along the $y$ direction. The two points
shown in each of the panels are the sites for which the $zz$ correlations
were computed.
}
\label{fig:systems-continuum-approximation}
\end{figure*}

We apply an additional approximation to the action $S_{\alpha}$
by writing the sum over the positions $z_j$ as an integral.
We replace the sum over positions $\sum_{j=1}^N \phifull{j}^2$
in Eq.~\eqref{eq:quadratic-action} by a term proportional
to the integral $\int_D  d z d\bar{z} \phi(z, \bar{z})^2$,
where $D$ is the region in the complex plane in which the spins are located.
The precise form of this replacement depends on the given system and is
provided in Appendix~\ref{sec:appendix-continuum-approximation},
where we also compute the approximate $zz$ correlations of
Eq.~\eqref{eq:zz-correlations-approximate-expression-path-integral} using
the Green's function formalism. We considered the systems
illustrated in Fig.~\ref{fig:systems-continuum-approximation}:
a 1D system (infinite line and circle), a 2D system without a boundary (sphere),
and a 2D system with a boundary (half plane and half-infinite cylinder).
Our results in the continuum limit are
summarized in Table~\ref{tab:summary-continuum-limit}.

Note that the
continuum approximation predicts a power law decay of the long-range
correlations in a 1D and at the edge of a 2D system with the power
being $-2$ independent of $\alpha$. In the bulk of a 2D system,
we find an exponential decay of the correlations at large distances.

\begin{table*}[htb]
\caption{\label{tab:summary-continuum-limit}
Approximation of the $zz$ correlations in $\psi_\alpha$ in the
continuum approximation, cf. Fig.~\ref{fig:systems-continuum-approximation}
for an illustration of the different systems.
$\Phi(z, s, a) = \sum_{m=0}^{\infty} z^m/(a + m)^s$ denotes
the Lerch transcendent function, $\Si$ and $\Ci$ are the sine and
cosine integral functions, respectively. The position on the sphere
$\spherepos{i} \in S^2$ is defined in
Eq.~\eqref{eq:definition-omega-sphere}.
}
\begin{ruledtabular}
\begin{tabular}{lll}
System & $zz$ correlations & Large-distance limit ($l \gg 1$) \\
\hline 
Infinite line &
$\zz{l+1}{1} \approx 2\alpha \left[\gaux{r}\right]$,&
$-\frac{1}{2 \pi^2 \alpha} \frac{1}{l^2}$
\\ 
($l = 0, 1, 2, \dots $) & with $r=2 \pi \alpha l$ & \\

&&\\

Circle & $\zz{l+1}{1} \approx \frac{1}{N} - 2 \alpha \mathrm{Re}\left[\Phi\left(e^{2 \pi i \frac{l}{N}}, 1, N \alpha\right)\right]$  &  \\ 
($l=0, 1, \dots , N-1$) && \\

&&\\

Sphere $\left(\spherepos{i}, \spherepos{j} \in S^2\right)$ &
$\zz{i}{j} \approx -\alpha \int_{0}^{\infty} d q \frac{2 \cos\left(\sqrt{N \alpha - \frac{1}{4}} q\right)}{\sqrt{2 \cosh(q) - 2 + |\spherepos{i} - \spherepos{j}|^2 }}$ &\\

&&\\

Half-plane &
$\zz{i_x, l+1}{i_x, 1} \approx \int_{-\infty}^\infty \frac{d q}{2 \pi} e^{-i q l} g(q)$, &
edge ($i_x = 1$): $-\frac{1}{\pi l^2}$ \\
$\begin{aligned}
(i_x &= 1, 2, \dots;\\
l &= 0, 1, 2, \dots)
\end{aligned}$
 &with $g(q) = -\frac{2 \pi  \alpha}{\sqrt{4 \pi  \alpha +q^2}}  \left[1 + \frac{4 \pi  \alpha  e^{-2 \left(i_x-1\right) \sqrt{4 \pi  \alpha +q^2}}}{\left(\left| q\right| +\sqrt{4 \pi  \alpha +q^2}\right)^2}\right]$ & bulk ($i_x \to \infty$): $ -\pi^{\frac{1}{4}} \alpha^{\frac{3}{4}} \frac{e^{-2 \sqrt{\pi \alpha} l}}{\sqrt{l}}$\\

&&\\ 

Half-infinite cylinder & $\zz{i_x, l+1}{i_x, 1} \approx \frac{1}{N_y} \sum_{m=-\infty}^{\infty} e^{-\frac{2 \pi i m l}{N_y}} g(\frac{2 \pi m}{N_y})$, &  \\
$\begin{aligned}
(i_x &= 1, 2, \dots;\\
l &= 0, 1, \dots, N_y - 1)
\end{aligned}$ &with $g(q)$ as for the half-plane&
\end{tabular} 
\end{ruledtabular}
\end{table*}

\subsection{Discrete approximation}
\label{sec:discrete-approximation}
The continuum approximation of the previous subsection does not take into
account the lattice structure of the states $\psi_{\alpha}$.
In order to test this simplification and to obtain a better approximation
of the exact $zz$ correlations, we now discuss an approximation scheme
that keeps the lattice. Specifically, we discuss three types of lattices:
a uniform lattice on a circle, an approximately uniform lattice on the
sphere, and a square lattice on the cylinder.
Thus, we can study both a 1D system and
2D systems with and without edges.

In the discrete case, we do not work with the path integral representation of
the approximate $zz$ correlations of Eq.~\eqref{eq:quadratic-action} since
it contains short-distance divergences at the lattice positions $z_j$.
Taking the continuum limit as done above is one way to remove
these divergences. In the discrete case, we choose to work with normal
ordered fields and thus avoid divergences.
This corresponds to taking the normal ordered expression of
Eq.~\eqref{eq:zz-correlations-exact-expression-before-path-integral}
as the starting point of the approximation instead of the path
integral representation of Eq.~\eqref{eq:zz-correlations-exact-expression-path-integral}.
The approximation of the $zz$ correlation then becomes
\begin{widetext}
\begin{align}
\zz{i}{j} &\approx - \alpha \frac{\langle :\phifull{i} e^{-\frac{\alpha}{2} \phifull{i}^2}: :\phifull{j} e^{-\frac{\alpha}{2} \phifull{j}^2}: \prod_{k (\neq i, j)}^N : e^{-\frac{\alpha}{2} \phifull{k}^2}: \rangle}{\langle \prod_{k=1}^N : e^{-\frac{\alpha}{2} \phifull{k}^2}:\rangle}.
\end{align}
\end{widetext}
In Appendix~\ref{sec:appendix-discrete-approximation}, we compute this expression and find
\begin{align}
\label{eq:approximation-in-discrete-case}
\zz{i}{j} &\approx \Gamma_{i, j},
\end{align}
where
\begin{align}
  \label{eq:Gamma}
  \Gamma_{i, j} &= \left[\Gammamat\right]_{i, j},
\end{align}
$r \in \{1, \dots, N\}$ is an arbitrary index, $M$ and $T_r$
are the $N \times N$ with entries
\begin{align}
  \label{eq:matrix-M}
  M_{m, n} &= \Mmat{m}{n},\\
  \left(T_r\right)_{m, n} &= \Trmat{m}{n},
\end{align}
$\evec{r}$ is the $r$th unit vector, and $\mathbb{I}$ is
the identity matrix. The $N \times N$ matrix $d_{i, j}$
contains the distances between sites $i$ and $j$.
It is given by $d_{i, j} = |z_i - z_j|$ for positions in the complex plane.
We note that the matrix $\Gamma$ is independent of the choice of $r$.

\subsubsection{Definition of lattices}
We now define the lattices on the circle, sphere, and the cylinder
and discuss how to ensure that the approximation retains
the symmetries of the lattice.

The positions of a uniform lattice on the circle are given
by
\begin{align}
  \label{eq:lattice-definition-circle}
  z_j &= e^{\frac{2 \pi i}{N} j}, \text{ with } j \in \{1, \dots, N\}.
\end{align}

In the case of the sphere, we would like to work with a lattice
that has an approximately uniform distribution of points on the
sphere embedded in three-dimensional space.
Following Ref.~\onlinecite{Nielsen2012},
we generate such a lattice by minimizing
the objective function
\begin{align}
  \label{eq:fobjective-lattice-sphere}
  \sum_{i < j}^N \frac{1}{|\spherepos{i} - \spherepos{j}|^2},
\end{align}
where $\spherepos{j}$ is given in terms of the polar angle $\theta_j$
and the azimuthal angle $\varphi_j$ as
\begin{align}
  \label{eq:definition-omega-sphere}
  \spherepos{j} = \left(\begin{matrix}
      \sin(\theta_j) \cos(\varphi_j)\\
      \sin(\theta_j) \sin(\varphi_j)\\
      \cos(\theta_j)
    \end{matrix}\right).
\end{align}
When computing the exact $zz$ correlations, the positions on the sphere
can be mapped to the complex plane
using the stereographic projection
$z_j = \tan\left(\theta_j/2\right) e^{-i \varphi_j}$.
For the approximate $zz$ correlations, however,
it is better not to do this projection but to work directly on the sphere.
The reason is that the differences in the complex plane,
$|z_m - z_n| = |e^{-i \varphi_m} \tan(\theta_m/2) - e^{-i \varphi_n} \tan(\theta_n/2)|$,
are not invariant under general rotations of the sphere.
Note that this is not a problem for the exact $zz$ correlations since
\begin{align}
\langle \sigma^z_i \sigma^z_j \rangle &= 
 \frac{\sum_{s_1, \dots, s_N} s_i s_j \delta_{\mathbf{s}} \prod_{m < n} |z_m - z_n|^{2 \alpha s_m s_n}}
 {\sum_{s_1, \dots, s_N} \delta_{\mathbf{s}} \prod_{m < n} |z_m - z_n|^{2 \alpha s_m s_n}} \notag \\
 \label{eq:rotation-invariance-of-exact-zz-correlations}
 & = \frac{\sum_{s_1, \dots, s_N} s_i s_j \delta_{\mathbf{s}} \prod_{m < n} |\spherepos{m} - \spherepos{n}|^{2 \alpha s_m s_n}}
  {\sum_{s_1, \dots, s_N} \delta_{\mathbf{s}} \prod_{m < n} |\spherepos{m}-\spherepos{n}|^{2 \alpha s_m s_n}},
\end{align}
where $|\spherepos{m} - \spherepos{n}|$ is invariant under sphere rotations.
[Eq.~\eqref{eq:rotation-invariance-of-exact-zz-correlations} follows from
$|\spherepos{m} - \spherepos{n}| = 2 \cos(\theta_m/2) \cos(\theta_n/2) |z_m - z_n|$
and $s_1 + \dots + s_N = 0$.]
As we show in Appendix~\eqref{sec:appendix-free-boson-sphere},
the replacement of $|z_m - z_n|$ by $|\spherepos{m} - \spherepos{n}|$ corresponds
to working directly on the sphere instead of the complex plane.
Doing our approximation for the free-boson field $\phi(\theta, \varphi)$
instead of $\phi(z, \bar{z})$ thus leads us to an approximation that
keeps the rotation invariance on the sphere.
It is given by the expressions following
Eq. \eqref{eq:approximation-in-discrete-case}
with $d_{m, n} = |\spherepos{m} - \spherepos{n}|$.

The positions on the cylinder with $N_x$ sites in the open direction
and $N_y$ sites in the periodical direction are defined as
\begin{align}
  \label{eq:lattice-definition-cylinder}
  w_j &= \frac{2 \pi}{N_y} (j_x + i j_y),
\end{align}
where $j_x \in \{1, \dots, N_x\}$ and $j_y \in \{1, \dots, N_y\}$
are the $x$ and $y$ components of the index $j$,
respectively [$j = (j_x - 1) N_y + j_y$].
The positions $w_{j_x, j_y + N_y}$ and $w_{j_x, j_y}$ are identified to
impose periodicity in the $y$ direction. Usually, the coordinates $w_j$
on the cylinder are projected onto the complex plane through $z_j = e^{w_j}$.
As for the stereographic projection in the case of the sphere, this mapping is,
however, not a symmetry of the approximate $zz$ correlations.
Even though $|z_i - z_j|$ does not change under rotations of the cylinder,
it distorts distances. As we show in Appendix~\ref{sec:appendix-free-boson-cylinder},
working with coordinates $w_j$ instead of $z_j = e^{w_j}$ corresponds to the
replacement $|z_i - z_j| \to d_{i, j}$ with
\begin{align}
\label{eq:distances-on-cylinder}
d_{i, j} &= \left|2 \sinh\left(\frac{1}{2} (w_i - w_j)\right)\right|.
\end{align}
Our approximation on the cylinder is thus obtained by expanding in the field
$\phi(w, \bar{w})$ instead of $\phi(z, \bar{z})$.
The resulting approximation of $\zz{i}{j}$
is given by using $d_{i, j}$ of Eq.~\eqref{eq:distances-on-cylinder} in
the expressions following Eq.~\eqref{eq:approximation-in-discrete-case}.

\subsubsection{Choice of the lattice scale}
\label{sec:choice-of-lattice-scale}
The exact $zz$ correlations
are invariant under rescaling transformations of the lattice due to
the conformal symmetry of the correlator of Eq.~\eqref{eq:definition-of-wave-function}.
These change the distances according to
\begin{align}
  \label{eq:rescalings-of-lattice}
  d_{m, n} \to \lambda d_{m, n},
\end{align}
where $\lambda > 0$.
The quadratic, discrete approximation of $\zz{i}{j}$ is, however,
not invariant under such rescalings since the matrix $M$ of
Eq.~\eqref{eq:matrix-M} varies under a change of the lattice scale $\lambda$.
Thus, different choices of $\lambda$ lead to different
values of the approximation and we need a criterion to uniquely determine the value
of $\lambda$. We note that this problem does not occur in the continuum approximation
since the replacement of the sum over lattice positions by an integral restores scale
invariance as shown in Appendix~\ref{sec:appendix-continuum-approximation}.
In order to fix $\lambda$ in the discrete case, we computed the subleading term
of the expansion for the $zz$ correlations in
Appendix~\ref{sec:appendix-discrete-approximation}:
\begin{align}
\label{eq:approximation-with-subleading-term}
\zz{i}{j} \approx \Gamma_{i,j} -  \Gamma_{i, j} (\Gamma_{i,i} + \Gamma_{j, j}).
\end{align}
For two given indices $i$ and $j$, we define the optimal scale $\lambda$
by requiring that it
minimizes the subleading term of the expansion, i.e. it is a minimum of
\begin{align}
\label{eq:objective-lambda-ij}
|\Gamma_{i,i} + \Gamma_{j, j}|.
\end{align}
Instead of choosing different scales for the different values of $i$ and $j$,
we also considered the following simpler approach:
For a given set of positions and value of $\alpha$, we determine a
single optimal scale $\lambda$ by minimizing the expression
\begin{align}
  \label{eq:objective-lambda}
  \sqrt{\sum_{j=1}^N \Gamma_{j, j}^2}.
\end{align}

In addition to minimizing expression~\eqref{eq:objective-lambda-ij}
or~\eqref{eq:objective-lambda}, we require that $\lambda$
is chosen such that
eigenvalues of the matrix $T_r^t M T_r + \mathbf{e}_r \mathbf{e}_r^t$
appearing in Eq.~\eqref{eq:Gamma} are all positive.
This condition ensures the convergence of an $N$-dimensional
Gaussian integral, cf. the derivation in
Appendix~\ref{sec:appendix-discrete-approximation}.
As our numerical calculations show, these requirements uniquely
determine the value of $\lambda$.

We did computations with multiple optimizations, i.e. minimizing
expression~\eqref{eq:objective-lambda-ij}, and a single
optimization of expression~\eqref{eq:objective-lambda}. We found
that doing multiple optimizations does not result in a substantial
improvement of our results.
Therefore, the data shown in the
following correspond to a single optimization for a given lattice
and value of $\alpha$.

\section{Quality of approximation for different systems}
\label{sec:quality-of-approximation}
In this Section, we test the validity of our approximation
by comparing the obtained $zz$ correlations to their actual value.
Our aim in doing this analysis is to test whether
the simple picture of the effective, quadratic theory
is accurate.
In two special cases, the actual $zz$ correlations can be computed exactly,
namely for $\alpha=\frac{1}{4}$,
where $\psi_{\alpha}$ is the wave function of $\frac{N}{2}$ free
fermions~\cite{Nielsen2012}, and for $\alpha=\frac{1}{2}$ on the circle,
where the exact correlations are known analytically~\cite{Nielsen2011, Stephan2017}.
For all other cases,
we used a Metropolis Monte Carlo method to obtain estimates of
the exact $zz$ correlations in $\psi_{\alpha}$. The approximation
data in the following plots
was computed using the discrete scheme described in
Sec.~\ref{sec:discrete-approximation}. We restrict ourselves to the
four values of $\alpha=0.125, 0.25, 0.375$, and $\alpha=0.5$ here.
Plots for additional values of $\alpha$ as well as for the
continuum approximation can be found in the Supplementary Material\cite{supplement}.

\subsection{One-dimensional system}
We first consider a 1D system. In this case, we can compare our
results to bosonization studies of the XXZ model
\begin{align}
H_{\mathrm{XXZ}} &= \sum_{j=1}^N \left(t^x_j t^x_{j+1} + t^y_j t^y_{j+1} + \Delta t^z_j t^z_{j+1} \right),
\end{align}
where $t^a_{N+1} = t^a_1$ for periodic boundary conditions. This model is in a critical
phase for anisotropies $-1 < \Delta \leq 1$.
The wave function $\psi_{\alpha}$ was used previously~\cite{Cirac2010}
as a variational ansatz for $H_{\mathrm{XXZ}}$ in the half-filled sector
($t^z_1 + \dots + t^z_N = 0$).
More precisely, the positions $z_j$ were taken to be uniformly distributed
on a circle and $\alpha$ was determined for a given value of
$\Delta$ such that the variational energy in $\psi_{\alpha}$ is minimal.
In the critical phase, the optimal value of $\alpha$ approximately satisfies
$\Delta = -\cos(2 \pi \alpha)$ and the overlap
between the exact ground state $\psi_0$ and the optimal $\psi_{\alpha}$
is large~\cite{Cirac2010}. Using the relation $\Delta = -\cos(2 \pi \alpha)$,
the results for the long-range $zz$ correlations obtained using
bosonization~\cite{Luther1975, Lukyanov1998} are given by
\begin{align}
\label{eq:corr-1D-from-bosonization}
\frac{\langle \psi_0 | \sigma^z_{l+1} \sigma^z_1 | \psi_0 \rangle}{\langle  \psi_0 | \psi_0  \rangle}
&\sim -\frac{1}{2 \pi^2 \alpha} \frac{1}{l^2} + A \frac{(-1)^l}{l^{\frac{1}{2 \alpha}}},
\end{align}
where $A$ is the $\alpha$-dependent amplitude that
was determined in Ref.~\onlinecite{Lukyanov1998}.
In the bosonization formalism,
the local Pauli matrix $\sigma^z_i$ has two contributions: A smooth term proportional
to the U(1) current $\partial_y \phi(y)$ and a second, rapidly varying term proportional
to $\cos(\pi y + \sqrt{2 \pi} \phi(y))$.\cite{Cabra2004}
The decay of correlations with a power of
$-2$ originates from the correlator of two U(1) currents, while the rapidly varying term
in the representation of $\sigma^z_i$ causes the staggered contribution to
Eq.~\eqref{eq:corr-1D-from-bosonization}.

For small values of $\alpha$, the first term in Eq.~\eqref{eq:corr-1D-from-bosonization}
is dominant. Our analytical results for the long-range
correlations in an infinite 1D system of
Table~\ref{tab:summary-continuum-limit} correctly reproduce this term.
As $\alpha$ approaches $\frac{1}{4}$, however, the second, alternating term
in Eq.~\eqref{eq:corr-1D-from-bosonization} becomes relevant and eventually
dominates for $\alpha > \frac{1}{4}$. In this regime, the state develops
quasi-long-range antiferromagnetic order.
This behavior is not captured by the results of our continuum approximation.
In contrast to bosonization, our representation of the $zz$ correlations
neglects rapid changes with position by assuming that the boson field is close to
the same extremum of cosine at all positions, cf.
Sec.~\ref{sec:effective-theory}. Furthermore, our approximation assumes that
$\sqrt{\alpha} \phi(z, \bar{z})$ is
small and therefore we do not expect to obtain the second term in
Eq.~\eqref{eq:corr-1D-from-bosonization}, which is relevant for large $\alpha$.

\begin{figure*}[htb]
  \centering
  \includegraphics{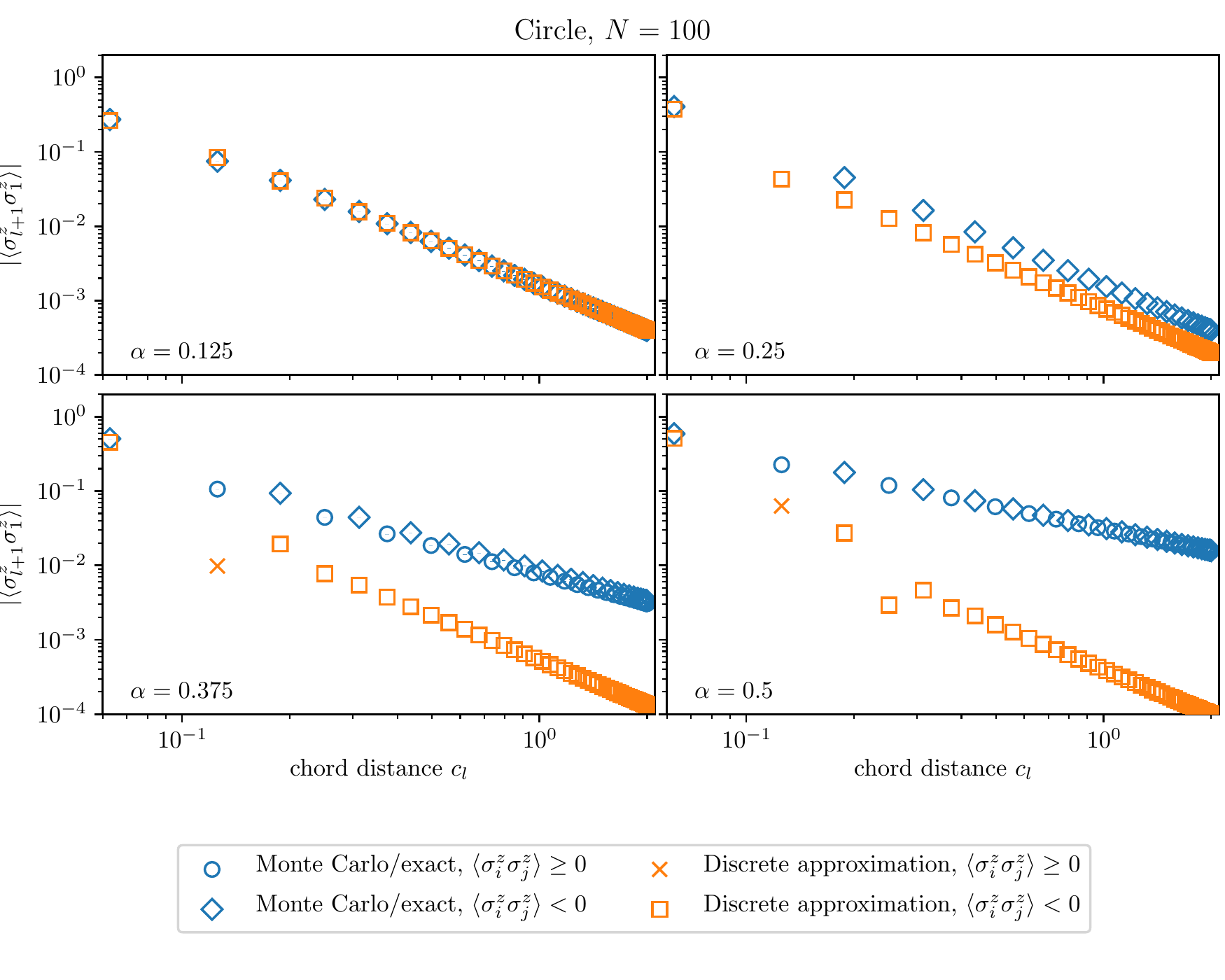}
  \caption{\label{fig:circle} (Color online) $zz$ correlations
  in $\psi_\alpha$ for $N=100$ spins uniformly distributed on the circle.
  The data shown in blue are Monte Carlo estimates
  ($\alpha=0.125$ and $\alpha=0.375$) or exact ($\alpha=0.25$ and $\alpha=0.5$).
  The Monte Carlo errors are of the order of $10^{-6}$ and are thus not visible.
  The horizontal axes show the chord distance $c_{l} = |2 \sin(\pi l/N)|$.
  At $\alpha = 0.25$, every second value of the actual correlator (blue symbols)
  is zero within numerical error and thus not visible.
  }
\end{figure*}
Let us now discuss our numerical results for the discrete approximation. These
are shown for $N=100$ spins uniformly distributed on the circle in Fig.~\ref{fig:circle}.
At large  distances, we observe a polynomial
decay in the correlations. (Both axes in Fig.~\ref{fig:circle} are scaled logarithmically.)
We find good agreement between the Monte Carlo estimate of the exact correlator and the
approximation for $\alpha = 0.125$. The results
differ substantially from the exact correlator for $\alpha \ge 0.25$ and the deviation
grows with $\alpha$. In particular, the oscillating behavior of $\zz{l+1}{1}$
for $\alpha > 0.25$ is not reproduced by our approximation. 

\begin{figure*}[htb]
\centering
  \includegraphics{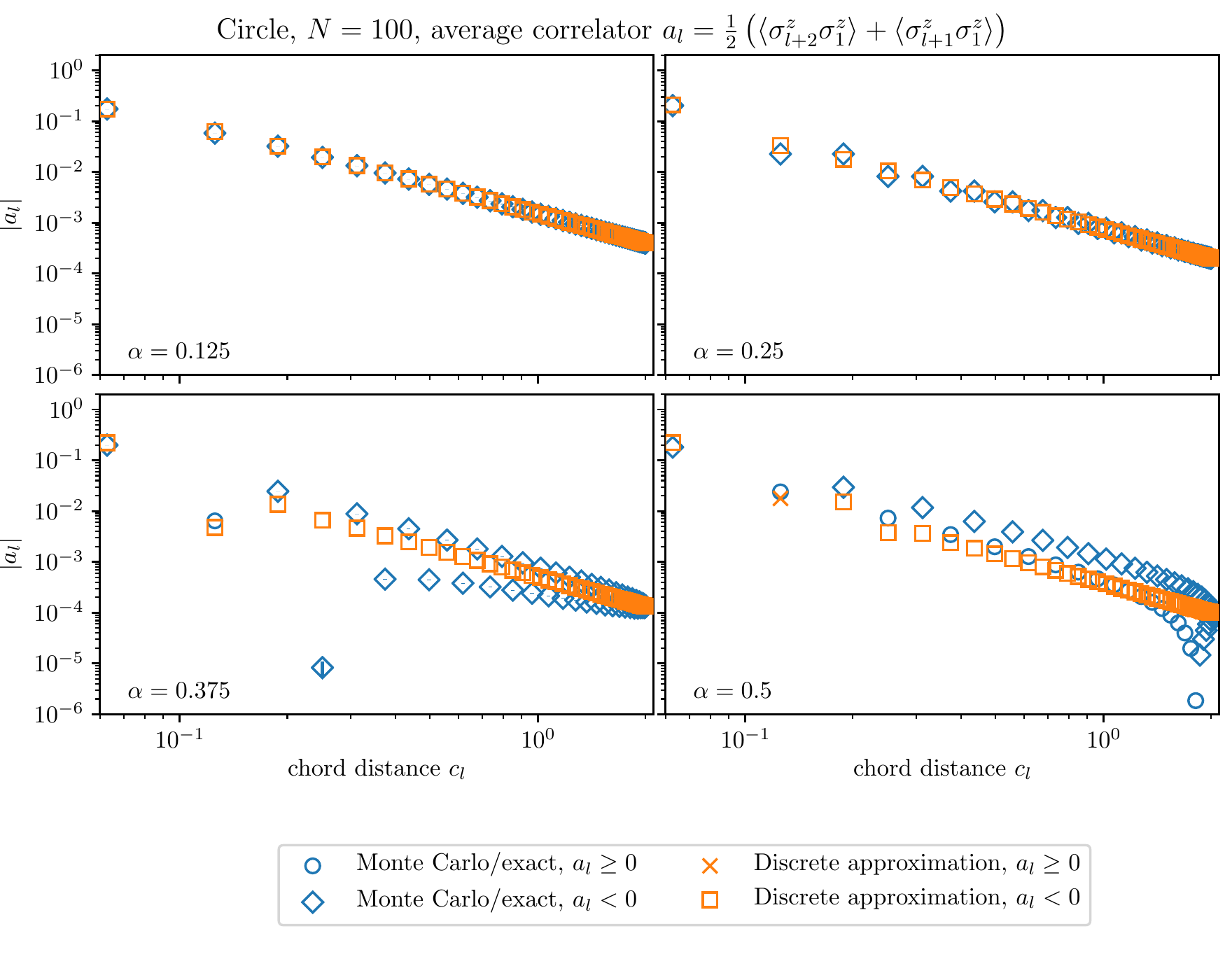}
\caption{\label{fig:circle2} (Color online)
Average $a_l = \aldef$  between
in the $zz$ correlations in $\psi_{\alpha}$ at distances
$l$ and $l+1$ for $N=100$ spins on the circle. The blue symbols correspond
to numerically exact data for $\alpha=0.25$ and $\alpha=0.5$
and to Monte Carlo estimates for $\alpha=0.125$
and $\alpha=0.375$.
The horizontal axes show the chord distance $c_{l} = |2 \sin(\pi l/N)|$.
}
\end{figure*}
To further emphasize that the approximation captures the smooth part
of the correlator but not the alternating one, we consider the average
$a_l = \aldef$ between the correlator at
distances $l+1$ and $l$. This combination suppresses the
oscillating term at large distances:
\begin{align}
\label{eq:average-correlator-from-bosonization}
&\frac{1}{2} \left(
\frac{\langle \psi_0 | \sigma^z_{l+2} \sigma^z_1 | \psi_0 \rangle}{\langle  \psi_0 | \psi_0  \rangle} + 
\frac{\langle \psi_0 | \sigma^z_{l+1} \sigma^z_1 | \psi_0 \rangle}{\langle  \psi_0 | \psi_0  \rangle}
\right)\notag \\
&\quad\sim
-\frac{1}{2 \pi^2 \alpha l^2} + A \frac{(-1)^l}{4 \alpha l^{1 + \frac{1}{2 \alpha}}}
\end{align}
for the correlator obtained through bosonization. The average $a_l$
is plotted in Fig.~\ref{fig:circle2} for the actual $zz$ correlations
in $\psi_{\alpha}$ and those obtained within the discrete approximation. 
Indeed, the agreement for $a_l$ is better also for larger values of $\alpha$.
[The value of $\alpha = \frac{1}{2}$ is special in the sense that $a_l$ still
oscillates as a function of the distance.  The reason is that
both terms in Eq.~\eqref{eq:average-correlator-from-bosonization} for
$\alpha=\frac{1}{2}$ decay with the same power and the amplitude of the
oscillating term dominates.
While this oscillation is absent in the approximation, the power
of the decay of $a_l$ agrees with the actual one.]

In summary, both the comparison to
bosonization and to the exact correlator lead to the conclusion that our approximation
is only valid for small values of $\alpha$ in 1D.

\subsection{Two-dimensional system without a boundary (sphere)}
\begin{figure*}[htb]
\centering
\includegraphics{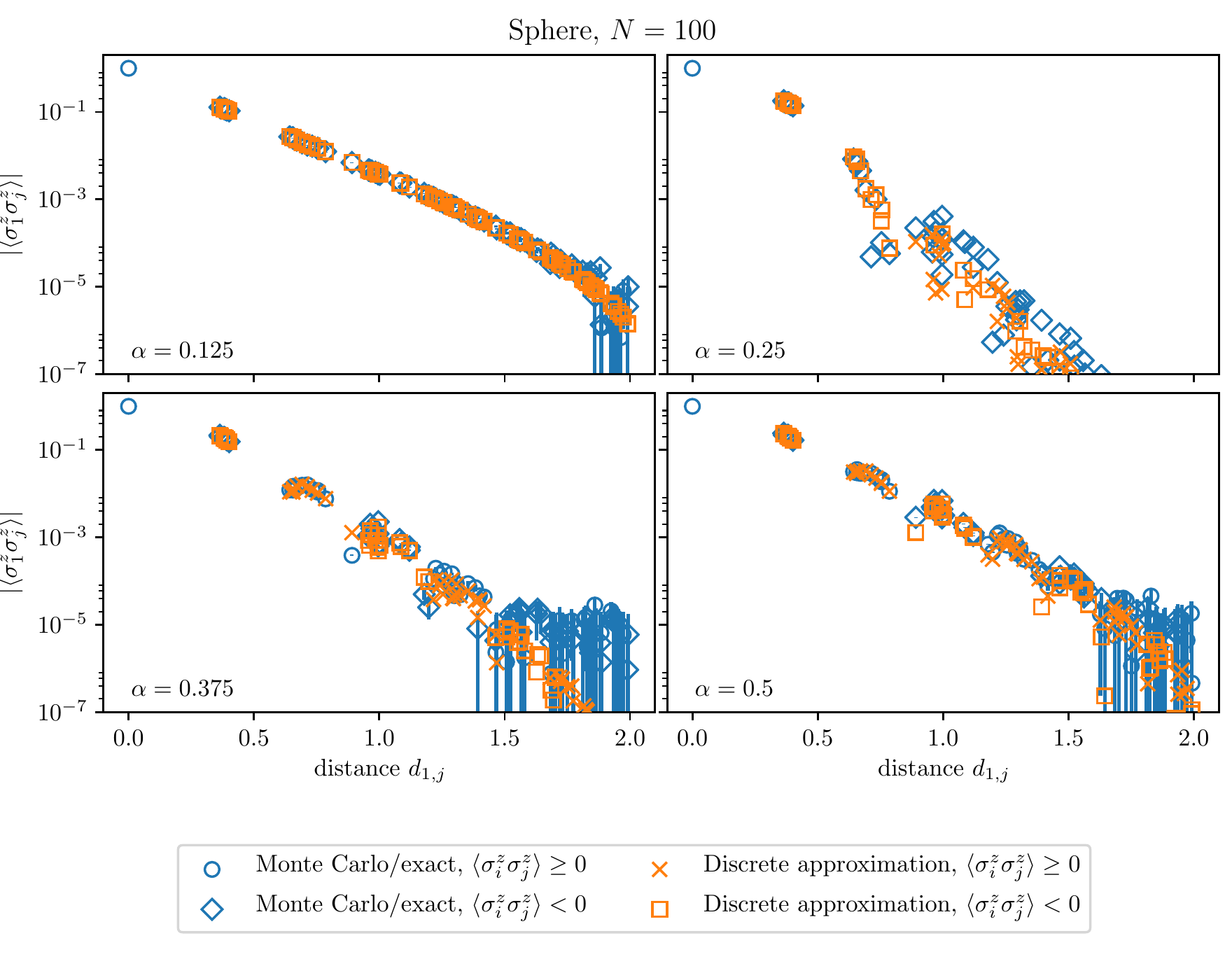}
\caption{\label{fig:sphere} (Color online) $zz$ correlations in $\psi_{\alpha}$
for $N=100$ spins with an approximately uniform distribution on the sphere.
The blue data points are Monte Carlo estimates ($\alpha=0.125, 0.375, 0.5$)
or exact ($\alpha=0.25$). The horizontal axes show the distance
$d_{1, j} = \left|\spherepos{1} - \spherepos{j}\right|$,
cf. Eq.~\eqref{eq:definition-omega-sphere} for the definition of $\spherepos{j}$.
}
\end{figure*}

We next discuss the case of a 2D system without a boundary. The results of
the discrete approximation are compared to the exact $zz$ correlations in
Fig.~\ref{fig:sphere} for $N=100$ spins on the sphere. The distribution
of sites on the sphere was chosen to be approximately uniform.
In Fig.~\ref{fig:sphere}, the vertical axis is scaled logarithmically, while
the horizontal axis is linear.

Both the approximate correlations and the actual ones decay exponentially.
Furthermore, a similar transition in the behavior of the $zz$ correlations
appears at $\alpha = \frac{1}{4}$ as in 1D: For $\alpha \le \frac{1}{4}$,
the correlations between distinct sites are
negative whereas for $\alpha > \frac{1}{4}$
they change sign as a function of the distance.
As in 1D, we find the best agreement
for the smallest value of $\alpha$ and large deviations at the transition
point $\alpha = \frac{1}{4}$.
However, in contrast to the 1D case, the approximation
captures the qualitative behavior of the correlations for larger values of
$\alpha$. In particular, we observe that the sign changes are reproduced correctly.

A reason for the better performance
of the approximation in 2D could be as follows:
Due to the cosine factors
in Eq.~\eqref{eq:zz-correlations-exact-expression-path-integral},
we assumed field configurations around $0$ mainly contribute
to the path integral. In the case of a 1D system, however,
these cosine factors only depend on the field configuration along
a 1D path and therefore they do not restrict contributions to the
path integral as strongly as for a 2D system.
Furthermore, the oscillations in the actual correlations are much stronger in 1D,
where they decay polynomially, than in 2D, where
the decay is exponential and there is no quasi-long-range
antiferromagnetic order.

\subsection{Two-dimensional system with a boundary (cylinder)}
\begin{figure*}[htb]
  \centering
  \includegraphics{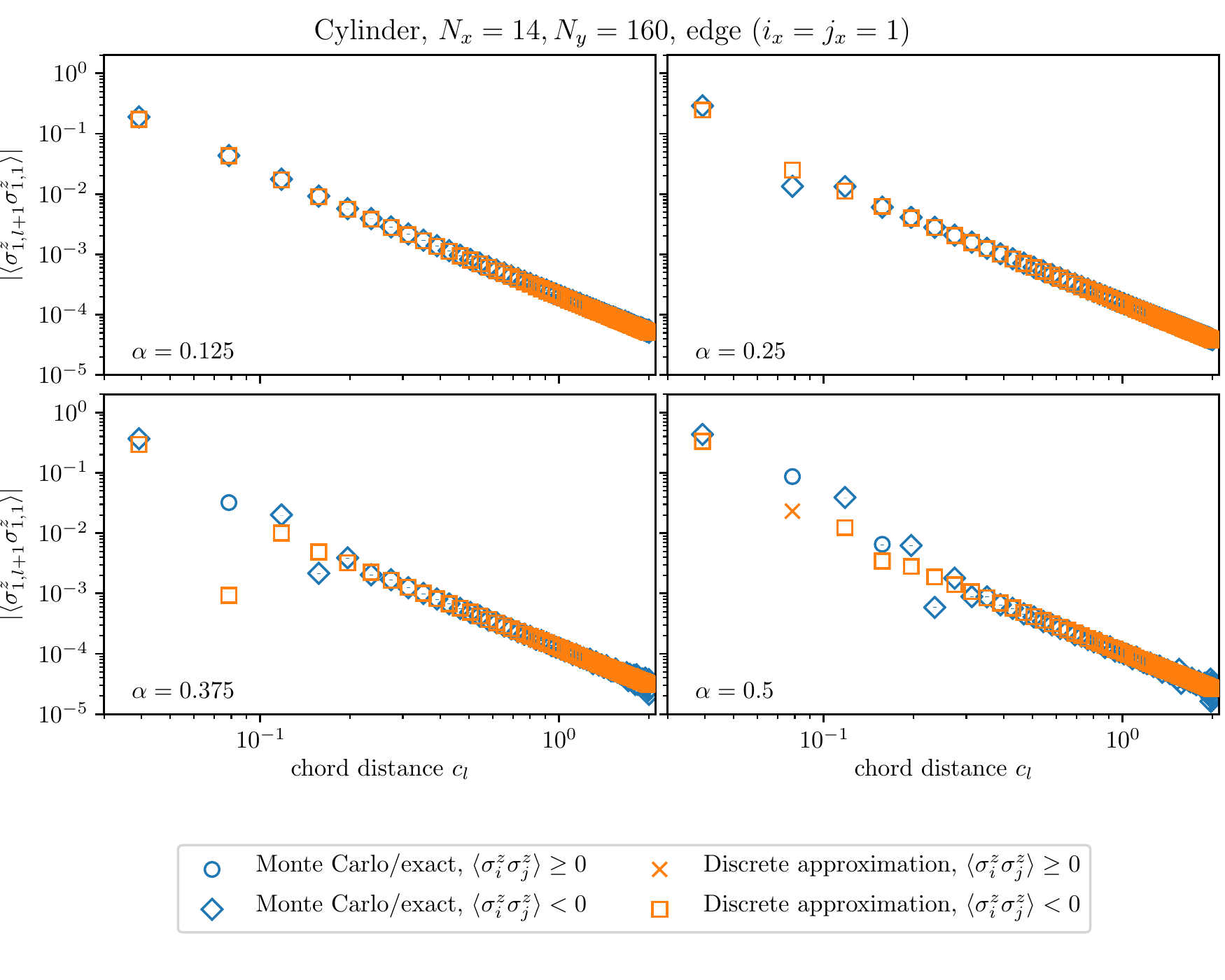}
  \caption{\label{fig:cylinder-edge} (Color online)
  $zz$ correlations in $\psi_{\alpha}$ on the edge of a cylinder with
  $N_x=14$ sites in the open direction and $N_y=160$ sites in the
  periodical direction. The shown correlations are those along the $y$
  direction. The blue data points are Monte Carlo estimates
  ($\alpha=0.125, 0.375, 0.5$) or exact ($\alpha=0.25$). The horizontal
  axes show the chord distance $c_l = |2 \sin(\pi l / N_y)|$.}
\end{figure*}

Let us now consider a 2D system with an edge. Our results
for the edge correlations of a cylinder of size $N_x = 14$ and $N_y=160$
are shown in Fig.~\ref{fig:cylinder-edge}. (The large number of spins
$N_y = 160$ is chosen so that we can
study the long-range decay of edge correlations.)
For $\alpha=0.125$, we observe a good agreement at all length scales.
As $\alpha$ becomes larger, the approximation deviates
from the exact result at short distances but follows
the decay at larger ones to good accuracy. The plots
in Fig~\ref{fig:cylinder-edge} are logarithmically scaled
on both axes and therefore show
an algebraic decay of long-range correlations with
a power of $-2$ as predicted
by our continuum approximation
(cf. Table~\ref{tab:summary-continuum-limit}).
This behavior corresponds to the decay of
a current-current correlator of a U(1) theory
at the edge.

\begin{figure*}[htb]
  \centering
  \includegraphics{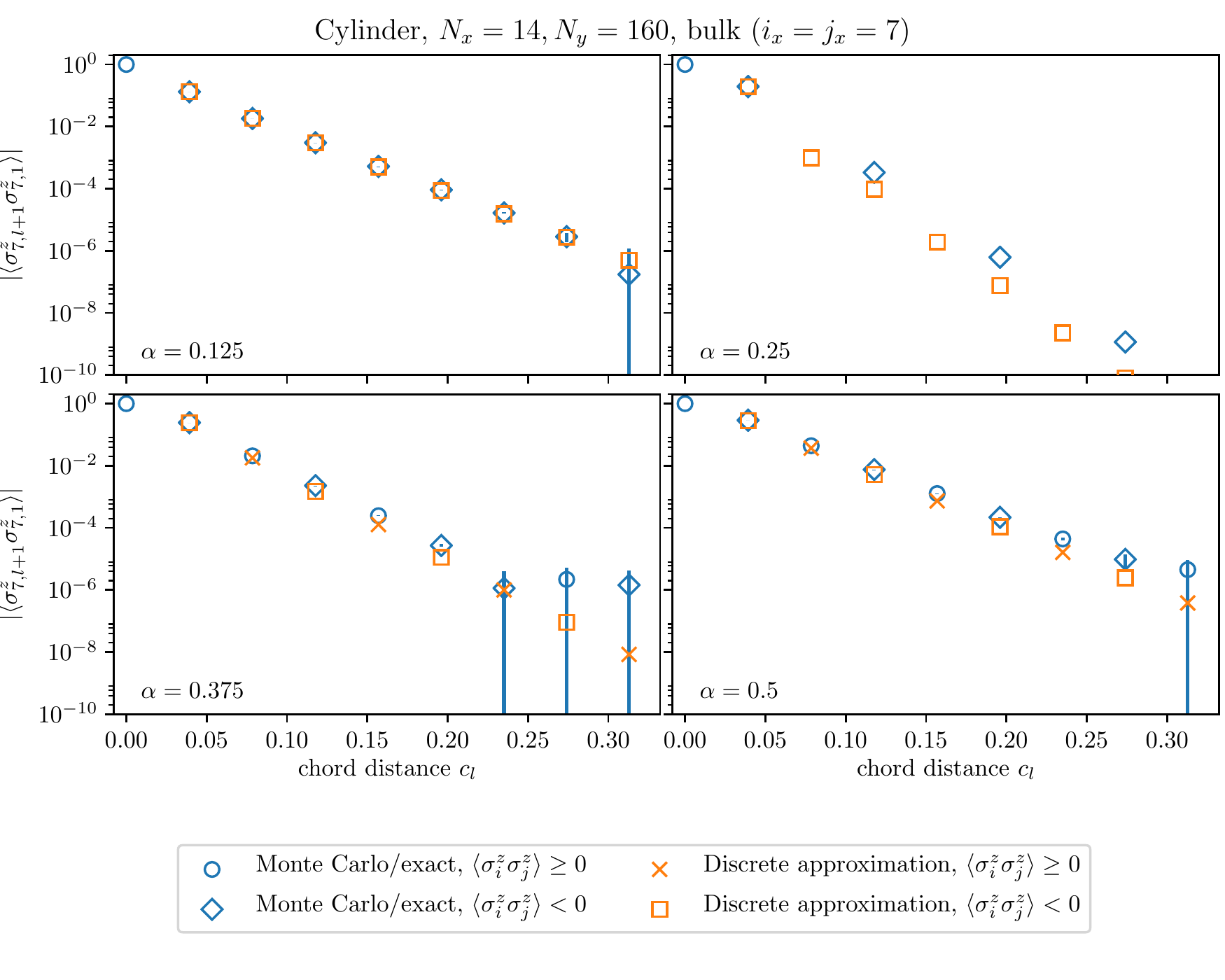}
  \caption{\label{fig:cylinder-bulk} (Color online)
  $zz$ correlations in $\psi_{\alpha}$ in the bulk of a cylinder with
  $N_x=14$ sites in the open direction and $N_y=160$ sites in the
  periodical direction. The shown correlations are those along the $y$
  direction. The blue data points are Monte Carlo estimates
  ($\alpha=0.125, 0.375, 0.5$) or exact ($\alpha=0.25$). The horizontal
  axes show the chord distance $c_l = |2 \sin(\pi l / N_y)|$.
  We only plot the data at short distances since the correlations decay
  exponentially and good Monte Carlo estimates cannot be obtained
  at larger distances.
  At $\alpha=0.25$, every second data point of the actual
  correlator (blue symbols)
  assumes a value below the lower boundary of the plot and is thus not visible.
  }
\end{figure*}

Our results for the correlator in the bulk of the cylinder are
shown in Fig.~\ref{fig:cylinder-bulk}.
The agreement between the approximate and the actual
correlator is similar to that on the sphere: The approximation
is best for $\alpha=0.125$, fails to describe the transition
at $\alpha=0.25$, and is qualitatively right for $\alpha > 0.25$.
The approximate correlations and the actual ones decay
exponentially. (The horizontal axes in Fig.~\ref{fig:cylinder-bulk}
are scaled linearly.)

\subsection{Comparison of continuum and discrete approximation}
Finally, we note that the discrete approximation generally yields better results than
the continuum approximation, especially for larger values of $\alpha$
in 2D.
(Corresponding plots are shown in the Supplementary Material\cite{supplement}.)
In particular, the continuum
approximation does not reproduce the alternating behavior of the correlations
for $\alpha > \frac{1}{4}$. At the edge of the 2D system, the continuum approximation
agrees with the exact correlation in the power of the long-range decay but
not in the prefactor, which is, however, reproduced correctly by the
discrete approximation.
In order to get a better agreement at larger values of $\alpha$,
it is therefore necessary to keep the lattice structure and to optimize the approximation
by choosing the scale as described in Sec.~\ref{sec:choice-of-lattice-scale}.

\section{Conclusion}
In this paper, we studied correlations in states $\psi_{\alpha}$ of $N$
spins on a lattice. The wave functions of $\psi_{\alpha}$ are
constructed
as chiral correlators of the CFT of a free, massless boson.
In 1D, they are approximate
ground states of the XXZ spin chain and they are similar to
Laughlin lattice states in 2D.

We derived an exact representation of the $zz$ correlations
in $\psi_{\alpha}$ in terms of a path integral expression.
By truncating it to quadratic order, we obtained
an effective description
in terms of a solvable, quadratic action.
Our effective theory differs from the original massless boson
by an additional mass-like term, which depends on
the field configuration at the lattice sites.
Thus, we establish an analytical connection between
physical properties of a state, namely its correlations,
and the underlying CFT.

By solving the free field theory, we obtained an
approximation of the $zz$ correlations in $\psi_{\alpha}$.
The mass-like term in our effective theory gives
rise to an exponential decay of correlations in
the bulk of a 2D system, whereas our approximation
predicts a power-law decay in 1D and at the edge
of a 2D system. We compared our results for the
approximate correlations to Monte Carlo estimates of the
exact value. Our analysis shows that the
approximation is most reliable for small values of $\alpha$,
where it yields good results in the case of a 1D
and a 2D system, both in the bulk and at the edge.
The long-range decay of the correlations at the edge
of a 2D system is reproduced correctly
by the approximation for all considered values of $\alpha$.
Furthermore, we find qualitative agreement between
the exact result and our approximating in the bulk
of a 2D system for values of $\alpha > \frac{1}{4}$,
whereas it fails to describe this regime in 1D.

The reason that the approximation is better in 2D than
in 1D is presumably due to the fact that field configurations
away from $0$ are more relevant in 1D than in 2D. We thus expect
to obtain the oscillating term for larger values of
$\alpha$ in 1D by taking into account these field
configurations.

In this work, we focused on approximating the $zz$ correlations
in terms of a free field theory. It would be interesting to
extend this method to other quantities.

\begin{acknowledgments}
This work was supported by
the Spanish government program FIS2015-69167-C2-1-P,
the Comunidad de Madrid grant QUITEMAD+ S2013/ICE-2801,
the grant SEV-2016-0597 of the Centro de Excelencia Severo Ochoa Programme,
the DFG within the Cluster of Excellence NIM,
the Carlsberg Foundation,
and the Villum Foundation.
\end{acknowledgments}

\begin{widetext}
\appendix
\section{Vertex operators and normal ordering}
\label{sec:appendix-normal-ordering}
In this section, we derive the following relation between
the exponential of the boson field $\phi(z, \bar{z})$ and the normal ordered exponential
of $\phi(z, \bar{z})$:
\begin{align}
  \label{eq:exponential-vs-normal-ordered-exponential}
  e^{i \sqrt{\alpha} s \phi(z, \bar{z})}
  &= e^{-\frac{1}{2} \alpha s^2 \langle \phi(z, \bar{z}) \phi(z, \bar{z}) \rangle} :e^{i \sqrt{\alpha} s \phi(z, \bar{z})}:,
\end{align}
where $\alpha > 0$ and $s \in \mathbb{R}$.

The field $\phi(z, \bar{z})$ satisfies Wick's theorem~\cite{DiFrancesco1997}, i.e.
$\phi(z, \bar{z})^n$ for $n \in \{0, 1, \dots\}$ is equal to the sum of all
possible contractions of $:\phi(z, \bar{z})^n:$. Any contraction of $k$ pairs of
fields from $:\phi(z, \bar{z})^n:$ results in the expression
\begin{align}
  \langle \phi(z, \bar{z}) \phi(z, \bar{z}) \rangle^k :\phi(z, \bar{z})^{n - 2k}:
\end{align}
and there are
\begin{align}
  \label{eq:appendix-combinatorial-factor}
  \frac{n (n-1) \dots (n-2k + 1)}{k! 2^k} = \frac{n!}{k! (n - 2k)! 2^k}
\end{align}
of such contractions.

The reason for this combinatorial factor is as follows: The numerator $n (n-1) \dots (n - 2k+1)$
counts the number of possibilities of taking $2k$ fields from $n$
fields. The factor $k!$ in the denominator corresponds to the
number of permutations of the factors $\langle \phi(z, \bar{z}) \phi(z, \bar{z}) \rangle^k$
and the factor $2^k$ corresponds to the freedom to interchange two fields within
a correlator $\langle \phi(z, \bar{z}) \phi(z, \bar{z}) \rangle$. Since these operators
do not change the contraction, we divide by $k! 2^k$.

Therefore,
\begin{align}
  \label{eq:appendix-phi-n}
   \phi(z, \bar{z})^n &=
   \sum_{k=0}^{[n/2]} \frac{n!}{k! (n - 2k)! 2^k} \langle \phi(z, \bar{z}) \phi(z, \bar{z}) \rangle^k :\phi(z, \bar{z})^{n - 2k}:,
\end{align}
where $[n/2] = n/2$ for $n$ even and $[n/2] = (n-1)/2$ for $n$ odd.
($[n/2]$ is the maximal number of pairs that can be contracted from $:\phi(z, \bar{z})^n:$.)
Multiplying Eq.~\eqref{eq:appendix-phi-n} by $(i \sqrt{\alpha} s)^n/n!$ and summing over $n$, we obtain
\begin{align}
  e^{i s \sqrt{\alpha} \phi(z, \bar{z})}
  &=\sum_{n=0}^{\infty}\sum_{k=0}^{[n/2]} \frac{1}{k! (n - 2k)! 2^k} (i \sqrt{\alpha} s)^n \langle \phi(z, \bar{z}) \phi(z, \bar{z}) \rangle^k :\phi(z, \bar{z})^{n - 2k}:\\
  &=\sum_{n=0}^{\infty}\sum_{k=0}^{[n/2]} \frac{\left(-\alpha s^2 \langle \phi(z, \bar{z}) \phi(z, \bar{z}) \rangle \right)^k}{k! 2^k} \frac{\left(i s \sqrt{\alpha} :\phi(z, \bar{z}):\right)^{n-2k}}{(n - 2k)!}.
\end{align}
This double sum can be rearranged into a double sum with both summation indices
ranging from $0$ to $\infty$ (cf. Ref.~\onlinecite{Arfken2001}),
\begin{align}
             e^{i s \sqrt{\alpha} \phi(z, \bar{z})}
  &= \sum_{n=0}^{\infty}\sum_{k=0}^{\infty} \frac{\left(-s^2 \alpha \langle \phi(z, \bar{z}) \phi(z, \bar{z}) \rangle \right)^k}{k! 2^k}
    \frac{\left(i s \sqrt{\alpha} :\phi(z, \bar{z}):\right)^{n}}{n!}
    \\
  &= e^{-\frac{1}{2} \alpha s^2 \langle \phi(z, \bar{z}) \phi(z, \bar{z}) \rangle} :e^{i \sqrt{\alpha} s \phi(z, \bar{z})}:.
\end{align}
It follows that
\begin{align}
\cos\left(\sqrt{\alpha} \phi(z, \bar{z})\right) &= e^{-\frac{1}{2} \alpha \langle \phi(z, \bar{z}) \phi(z, \bar{z}) \rangle} :\cos\left(\sqrt{\alpha} \phi(z, \bar{z})\right):
\intertext{and}
\sin\left(\sqrt{\alpha} \phi(z, \bar{z})\right) &= e^{-\frac{1}{2} \alpha \langle \phi(z, \bar{z}) \phi(z, \bar{z}) \rangle} :\sin\left(\sqrt{\alpha} \phi(z, \bar{z})\right):.
\end{align}
Therefore, the normal ordering can be left out in the exact
expression for the $zz$ correlations of
Eq.~\eqref{eq:zz-correlations-exact-expression-before-path-integral}.

\section{Free boson on the sphere and on the cylinder}
\label{sec:appendix-free-boson-sphere-cylidner}
In this section, we consider a free boson on the sphere and on the cylinder.
We explicitly work with coordinates $\Omega = (\theta, \varphi)$ on the
sphere and $(w, \bar{w})$ on the cylinder and do not project onto the plane.
This is not necessary when computing the exact correlations in $\psi_\alpha$,
which are invariant under a projection of the positions onto the complex plane.
However, different approximations of the correlations are obtained depending
on whether one expands in the field $\phi(z, \bar{z})$ or $\phi(\Omega)$ and
$\phi(w, \bar{w})$, respectively.
\subsection{Sphere}
\label{sec:appendix-free-boson-sphere}
The free boson on the sphere has the action
\begin{align}
  \label{eq:action-free-boson-on-sphere}
  S_{\mathrm{sphere}}[\phi]
  = \frac{1}{8 \pi} \int_{S^2} d \Omega
  \left[(\partial_{\theta}\phi(\Omega))^2 + \frac{1}{\sin^2(\theta)} (\partial_{\varphi} \phi(\Omega))^2
  + \tilde{m}^2 \phi(\Omega)^2 \right],
\end{align}
where $\Omega = (\theta, \varphi)$ in terms of the polar angle $\theta$ and the azimuthal angle $\varphi$,
$d \Omega = d \theta d \varphi \sin(\theta)$, and $\tilde{m}$ is a mass regulator. In this
section, we will take $\tilde{m}$ to be $0$ eventually.

After two integrations by part with vanishing boundary terms,
\begin{align}
  S_{\mathrm{sphere}}
  = \frac{1}{8 \pi} \int_{S^2} d \Omega \phi(\Omega) \left[L^2 + \tilde{m}^2 \right] \phi(\Omega),
\end{align}
where
\begin{align}
  L^2 &= -\frac{1}{\sin(\theta)} \partial_{\theta}
        \left[\sin(\theta) \partial_{\theta}\right] - \frac{1}{\sin^2(\theta)} \partial_{\varphi}^2
\end{align}
is the square of the orbital angular momentum operator ($L^2 = L_x^2 + L_y^2 + L_z^2$).

The two-point correlator $\langle \phi(\Omega) \phi(\Omega') \rangle$ is
determined by
\begin{align}
  \label{eq:greens-function-on-sphere-equation}
  \frac{1}{4 \pi} \left(L^2 + \tilde{m}^2 \right) \langle \phi(\Omega) \phi(\Omega') \rangle
  &= \delta(\Omega - \Omega'),
\end{align}
where the $\delta$ function is defined with respect to the measure $d \Omega$ [$\int_{S^2} d\Omega \delta(\Omega) = 1$].

Since the spherical harmonics $Y^{m}_l$ are eigenfunctions of $L^2$ with eigenvalues $l (l+1)$,
we expand
\begin{align}
  \langle \phi(\Omega) \phi(\Omega') \rangle &= \sum_{l, l'=0}^{\infty} \sum_{m=-l}^{l} \sum_{m'=-l'}^{l'} c^{m, m'}_{l, l'} Y^{m}_l(\Omega) Y^{m'*}_{l'}(\Omega').
\end{align}
Using $\delta(\Omega - \Omega') = \sum_{l=0}^{\infty} \sum_{m=-l}^l Y^{m}_l(\Omega) Y^{m*}_l(\Omega')$
and Eq.~\eqref{eq:greens-function-on-sphere-equation}, we obtain
\begin{align}
  c^{m, m'}_{l, l'} &= \frac{4 \pi}{l (l+1) + \tilde{m}^2} \delta_{l, l'} \delta_{m, m'},\\
  \langle \phi(\Omega) \phi(\Omega') \rangle &= \sum_{l=0}^{\infty} \frac{4 \pi}{l (l+1) + \tilde{m}^2}  \sum_{m=-l}^{l} Y^{m}_l(\Omega) Y^{m*}_l(\Omega').
\end{align}
The sum over $m$ is given by~\cite{Arfken2001}
\begin{align}
  \sum_{m=-l}^{l} Y^{m}_l(\Omega) Y^{m*}_l(\Omega') =  \frac{2 l + 1}{4 \pi} P_l(x),
\end{align}
where $P_l$ is the $l$th Legendre polynomial, $x = 1 - \frac{1}{2} |\mathbf{n}_{\Omega} - \mathbf{n}_{\Omega'}|^2$,
and $\mathbf{n}_{\Omega}$ and $\mathbf{n}_{\Omega'}$ are unit vectors on $S^2$ embedded in $\mathbb{R}^3$
[cf. Eq.~\eqref{eq:definition-omega-sphere}].

Let us first consider the case of a finite mass $\tilde{m} > 0$, which
is needed in Appendix~\ref{sec:appendix-condintuum-approximation-sphere}
for the computation of the continuum approximation on the sphere:
\begin{align}
\label{eq:greens-function-sphere-massive-intermediate}
\langle \phi(\Omega) \phi(\Omega') \rangle &= \sum_{l=0}^{\infty} \frac{2 l + 1}{l (l+1) + \tilde{m}^2} P_l(x)
= \sum_{l=0}^{\infty} \left(\frac{1}{l+\frac{1}{2} + \sqrt{\frac{1}{4} - \tilde{m}^2}} + \frac{1}{1 + \frac{1}{2} - \sqrt{\frac{1}{4} - \tilde{m}^2}}\right) P_l(x).
\end{align}
Using the generating function
of $P_l(x)$,
\begin{align}
  \label{eq:legendre-generating-function}
  \frac{1}{\sqrt{1 - 2 x z + z^2}} &= \sum_{l=0}^{\infty} P_l(x) z^l,
\end{align}
Eq.~\eqref{eq:greens-function-sphere-massive-intermediate} can be transformed
into an integral. To this end, we multiply the
generating function~\eqref{eq:legendre-generating-function} by $z^{\gamma-1}$
with $\mathrm{Re}(\gamma) > 0$ and then integrate $z$ from $0$ to $1$:
\begin{align}
\int_0^1 d z \frac{z^{\gamma - 1}}{\sqrt{1 - 2 x z + z^2}} &= \sum_{l=0}^{\infty} \frac{1}{l + \gamma} P_l(x).
\end{align}
Applying this identity to Eq.~\eqref{eq:greens-function-sphere-massive-intermediate},
we find
\begin{align}
\label{eq:greens-function-sphere-massive}
\langle \phi(\Omega) \phi(\Omega') \rangle &=
\int_0^1 d z
\frac{2 \cosh\left[\sqrt{\frac{1}{4} - \tilde{m}^2} \ln(z)\right]}
{\sqrt{z \left[(z - 1)^2 + |\mathbf{n}_{\Omega} - \mathbf{n}_{\Omega'}|^2 z \right]}}\\
&= \int_{0}^{\infty} d q \frac{2 \cosh\left(\sqrt{\frac{1}{4} - \tilde{m}^2} q\right)}{\sqrt{2 \cosh(q) - 2 + |\mathbf{n}_{\Omega} - \mathbf{n}_{\Omega'}|^2 }},
\end{align}
where $q = -\ln z$ was substituted in the integral. 

Let us now consider the case of $\tilde{m} \to 0$.
Since the $l=0$ term is divergent for $\tilde{m} \to 0$, we separate it from
the sum of Eq.~\eqref{eq:greens-function-sphere-massive-intermediate} and expand the remaining terms to
leading order in $\tilde{m}$:
\begin{align}
  \label{eq:greens-function-on-sphere-intermediate}
  \langle \phi(\Omega) \phi(\Omega') \rangle
  &= \frac{1}{\tilde{m}^2} + \sum_{l=1}^{\infty} \left(\frac{1}{l} + \frac{1}{l+1}\right) P_l(x) + \mathcal{O}(\tilde{m}^2),
\end{align}
where $(2 l + 1)/(l (l+1)) = 1/l + 1/(l+1)$ and $P_0(x) = 1$ was used.

As a last step, we apply the identities
\begin{align}
  \label{eq:legendre-sum1}
  \sum_{l=0}^{\infty} \frac{P_l(x)}{l + 1} &= \ln\frac{1 + \sqrt{2 - 2x} - x}{1-x},\\
  \label{eq:legendre-sum2}
  \sum_{l=1}^{\infty} \frac{P_l(x)}{l} &= \ln\frac{2}{1 + \sqrt{2 - 2x} - x},
\end{align}
which can be derived from the generating function of $P_l(x)$:
Eq.~\eqref{eq:legendre-sum1} follows from Eq.~\eqref{eq:legendre-generating-function}
by integrating $z$ from $0$ to $1$. To obtain Eq.~\eqref{eq:legendre-sum2}, one
can first bring the $l=0$ term in Eq.~\eqref{eq:legendre-generating-function}
to the left-hand side, then divide by $z$, and finally integrate $z$ from $0$ to $1$.

Using Eqs.~(\ref{eq:legendre-sum1}, \ref{eq:legendre-sum2}) in Eq.~\eqref{eq:greens-function-on-sphere-intermediate},
we obtain
\begin{align}
  \langle \phi(\Omega) \phi(\Omega') \rangle
  &= \frac{1}{\tilde{m}^2} - 1 + \ln\frac{4}{|\mathbf{n}_{\Omega} - \mathbf{n}_{\Omega'}|^2} + \mathcal{O}(\tilde{m}^2).
\end{align}
Taking $\tilde{m} \to 0$,
\begin{align}
  \langle \phi(\Omega) \phi(\Omega') \rangle &= -\ln |\mathbf{n}_{\Omega} - \mathbf{n}_{\Omega'}|^2 + \text{constant}.
\end{align}

Next, we define vertex operators on the sphere as
\begin{align}
  :e^{i \sqrt{\alpha} s \phi(\Omega)}:,
\end{align}
where $\alpha > 0$, $s \in \mathbb{R}$, and normal ordering is defined by
subtracting vacuum expectation values on the sphere:
\begin{align}
  :\phi(\Omega_1) \phi(\Omega_2):
  &= \phi(\Omega_1) \phi(\Omega_2) - \langle \phi(\Omega_1) \phi(\Omega_2) \rangle,
\end{align}
and similarly for more fields.

Let us now relate the vertex operator on the sphere to the
vertex operator on the plane. The bosonic field on the plane is
related to that on the sphere through the stereographic projection:
\begin{align}
  \phi(z, \bar{z}) &= \phi(\Omega),
\end{align}
with $z = \tan(\theta/2) e^{-i \varphi}$ and $\bar{z} = \tan(\theta/2) e^{i \varphi}$.
Correspondingly, we have
\begin{align}
  e^{i \sqrt{\alpha} s \phi(z, \bar{z})} &= e^{i \sqrt{\alpha} s \phi(\Omega)}.
\end{align}
Note that this equation only holds without normal ordering
since the normal ordering prescription depends on subtracting vacuum expectation
values on the plane and sphere, respectively. However, we can use
Eq.~\eqref{eq:exponential-vs-normal-ordered-exponential} relating
the exponential of $\phi$ to the normal ordered exponential of $\phi$.
[The computation leading to Eq.~\eqref{eq:exponential-vs-normal-ordered-exponential} was
formulated on the plane but it also applies to the sphere since
we only made use of the relation between normal ordering and subtractions
of vacuum expectation values.] Therefore,
\begin{align}
  e^{-\frac{1}{2} \alpha s^2 \langle \phi(\Omega) \phi(\Omega) \rangle} :e^{i \sqrt{\alpha} s \phi(\Omega)}:
  &= e^{-\frac{1}{2} \alpha s^2 \langle \phi(z, \bar{z}) \phi(z, \bar{z}) \rangle} :e^{i \sqrt{\alpha} s \phi(z, \bar{z})}:
\end{align}
or
\begin{align}
  :e^{i \sqrt{\alpha} s \phi(\Omega)}: &=
  :e^{i \sqrt{\alpha} s \phi(z, \bar{z})}: \lim_{\substack{\Omega' \to \Omega\\z' \to z}} e^{-\frac{1}{2} \alpha s^2 \left(\langle \phi(z, \bar{z}) \phi(z', \bar{z'}) \rangle - \langle \phi(\Omega) \phi(\Omega') \rangle\right)}\\
  &= :e^{i \sqrt{\alpha} s \phi(z, \bar{z})}: \lim_{\substack{\Omega' \to \Omega\\z' \to z}} \left(\frac{|z - z'|}{|\mathbf{n}_{\Omega} - \mathbf{n}_{\Omega'}|}\right)^{\alpha s^2}\\
  &=  :e^{i \sqrt{\alpha} s \phi(z, \bar{z})}: \left[2 \cos\left(\frac{\theta}{2}\right)^2\right]^{-\alpha s^2}.
\end{align}
In the last step, it was used that
\begin{align}
  \left|\mathbf{n}_{\Omega} - \mathbf{n}_{\Omega'}\right| &= 2 \cos\left(\frac{\theta}{2}\right) \cos\left(\frac{\theta'}{2}\right) \left|z - z'\right|.
\end{align}
We can now determine the correlator of $N$ vertex operators
on the sphere from the corresponding correlator on the plane:
\begin{align}
  \langle :e^{i \sqrt{\alpha} s_1 \phi(\Omega_1)} \dots :e^{i \sqrt{\alpha} s_N \phi(\Omega_N)}: \rangle
  &= \delta_{\mathbf{s}} \left(\prod_{m=1}^N  \left[2 \cos\left(\frac{\theta_m}{2}\right)^2\right]^{-\alpha s_m^2}\right) \prod_{m < n}^N |z_m - z_n|^{2 \alpha s_m s_n}\\
  &= \delta_{\mathbf{s}} \prod_{m < n}^N \left[2 \cos\left(\frac{\theta_m}{2}\right) \cos\left(\frac{\theta_n}{2}\right) |z_m - z_n|\right]^{2 \alpha s_m s_n}\\
  &= \delta_{\mathbf{s}} \prod_{m < n}^N |\mathbf{n}_{\Omega_m} - \mathbf{n}_{\Omega_n}|^{2 \alpha s_m s_n},
\end{align}
where we used that
\begin{align}
  \delta_{\mathbf{s}} \prod_{m < n}^N \left[2 \cos\left(\frac{\theta_m}{2}\right) \cos\left(\frac{\theta_n}{2}\right)\right]^{2 \alpha s_m s_n}
  &=   \delta_{\mathbf{s}} e^{\alpha \sum_{m \neq n} s_m s_n \ln\left[2 \cos(\theta_m / 2) \cos(\theta_n/2)\right]}\\
  &=   \delta_{\mathbf{s}} e^{-\alpha \sum_{m=1}^N s_m^2 \ln\left[2 \cos(\theta_m/2)^2\right]} \\
  &= \delta_{\mathbf{s}} \prod_{m=1}^N \left[2 \cos(\theta_m/2)^2\right]^{-\alpha s_m^2}.
\end{align}

\subsection{Cylinder}
\label{sec:appendix-free-boson-cylinder}
In the case of the cylinder, we can use the conformal transformation $z = e^w$ to transform the known correlator of vertex operators on the plane to that on the cylinder:
\begin{align}
\langle :e^{i \sqrt{\alpha} s_1 \phi(w_1, \bar{w}_1)}: \dots  :e^{i \sqrt{\alpha} s_N \phi(w_N, \bar{w}_N)}: \rangle &=  \left|\prod_{j=1}^N e^{\frac{\alpha}{2} w_j s_j^2}\right|^2 \langle :e^{i \sqrt{\alpha} s_1 \phifull{1}}: \dots  :e^{i \sqrt{\alpha} s_N \phifull{N}}: \rangle\\
&= \delta_{\mathbf{s}} \left|\prod_{j=1}^N e^{\frac{\alpha}{2} w_j s_j^2}\right|^2  \prod_{i < j}^N \left|e^{w_i} - e^{w_j}\right|^{2 \alpha s_i s_j}\\
&= \delta_{\mathbf{s}}   \prod_{i < j}^N \left|2 \sinh\left(\frac{1}{2} (w_i - w_j)\right)\right|^{2 \alpha s_i s_j}.
\end{align}
In the last step, it was used that
\begin{align}
\delta_{\mathbf{s}} \left|\prod_{j=1}^N e^{\frac{\alpha}{2} w_j s_j^2}\right|^2 &= \delta_{\mathbf{s}} \left|\prod_{i < j}^N e^{-\frac{\alpha}{2} s_i s_j (w_i + w_j)} \right|^2.
\end{align}

\section{Solution of the quadratic theory}
\label{sec:appendix-solution-quadratic-theory}

\subsection{Continuum approximation}
\label{sec:appendix-continuum-approximation}
The action determining the $zz$ correlations to quadratic order is given by
\begin{align}
\label{eq:quadratic-action-with-mass-regulator}
S_\alpha &= \frac{1}{8 \pi} \int d x d y\; \phi(x, y) (-\partial_x^2 - \partial_y^2 + \tilde{m}^2) \phi(x, y) + \frac{\alpha}{2} \sum_{j=1}^N \phi(x_j, y_j)^2,
\end{align}
where $\tilde{m}$ is a mass regulator. (We will take the limit $\tilde{m} \to 0$ eventually.)
In this section, we consider a continuum limit,
in which the term in the action that is proportional to $\alpha$ becomes an integral.

\subsubsection{One-dimensional system}
Let us first consider an infinite, one-dimensional lattice given by
\begin{align}
\label{eq:lattice-positions-1D-before-continuum-limit}
x_n = 0, y_n = 2 \pi \lambda n,
\end{align}
where $n \in \mathbb{Z}$ and $2 \pi \lambda > 0$ is the lattice constant.
We will later see that the resulting approximation for the $zz$ correlations
is independent of $\lambda$.

The term in the action
that is proportional to $\alpha$ becomes
\begin{align}
\label{eq:mass-term-1D-before-continuum-limit}
\frac{\alpha}{4 \pi \lambda} \sum_{n=-\infty}^{\infty} 2 \pi \lambda \phi(0, 2 \pi \lambda n)^2.
\end{align}
Taking the continuum approximation of this sum,
\begin{align}
\frac{\alpha}{4 \pi \lambda} \int_{-\infty}^{\infty} dy \phi(0, y)^2,
\end{align}
the action becomes
\begin{align}
S_\alpha &= \frac{1}{8 \pi} \int dx dy \; \phi(x, y) \left[-\partial_x^2 -\partial_y^2 + \tilde{m}^2 + 2 \frac{\alpha}{\lambda}\delta(x)\right] \phi(x, y)
\end{align}
The corresponding Green's function is determined by
\begin{align}
\label{eq:Greens-function-defining-equation-real-space-1D}
\left[-\partial_x^2 - \partial_y^2 + \tilde{m}^2 + 2 \frac{\alpha}{\lambda} \delta(x)\right] \Galpha(x, y; x', y') = 4 \pi \delta(x - x') \delta(y - y')
\end{align}
and provides an approximation of the $zz$ correlations through
\begin{align}
\label{eq:approximation-for-zz-1D}
\zz{m}{n} &\approx -\alpha \Galpha(0, 2 \pi \lambda m; 0, 2 \pi \lambda n).
\end{align}
Next, we go to Fourier space by expanding
\begin{align}
\Galpha(x, y; x', y') &= \int \frac{d p d q d p'}{(2 \pi)^3} e^{-i p x  - i p' x' - i q (y - y')} \Galpha(p, q; p').
\end{align}
This ansatz already incorporates the translational symmetry in the $y$ direction.
In Fourier space, Eq.~\eqref{eq:Greens-function-defining-equation-real-space-1D} becomes
\begin{align}
\label{eq:Greens-function-in-Fourier-space}
(p^2 + q^2 + \tilde{m}^2) \Galpha(p, q; p') + \frac{\alpha}{\lambda} \int \frac{d k}{\pi} \Galpha(k, q; p') &= 8 \pi^2 \delta(p + p').
\end{align}
Defining $F(p, q; p')$ through
\begin{align}
\label{eq:definition-of-F-from-G}
\Galpha(p, q; p') &= \frac{8 \pi^2}{p^2 + q^2 + \tilde{m}^2} \delta(p + p') + \frac{4 \pi F(p, q; p')}{(p^2 + q^2 + \tilde{m}^2) (p'^2 + q^2 + \tilde{m}^2)}
\end{align}
and multiplying Eq.~\eqref{eq:Greens-function-in-Fourier-space} by $p'^2 + q^2 + \tilde{m}^2$, we obtain
\begin{align}
F(p, q; p') + 2 \frac{\alpha}{\lambda} + \frac{\alpha}{\lambda} \int \frac{d k}{\pi} \frac{F(k, q; p')}{k^2 + q^2 + \tilde{m}^2} = 0.
\end{align}
From this equation, we conclude that $F(p, q; p')$ does not depend on $p$. Likewise, it does not depend on $p'$
since $F(p, q; p') = F(p', q; p)$, which follows from $\Galpha(x, y; x', y') = \Galpha(x', y; x, y')$. Therefore,
it follows that
\begin{align}
F(p, q; p') &= - 2 \left(\frac{\lambda}{\alpha} + \int \frac{d k}{\pi} \frac{1}{k^2 + q^2 + \tilde{m}^2}\right)^{-1}\\
&= -2 \frac{\sqrt{q^2 + \tilde{m}^2}}{1 + \frac{\lambda}{\alpha} \sqrt{q^2 + \tilde{m}^2}}
\intertext{and}
\Galpha(p, q; p') &=  \frac{8 \pi^2}{p^2 + q^2 + \tilde{m}^2} \delta(p + p') - 8 \pi \frac{\sqrt{q^2 + \tilde{m}^2}}{(p^2 + q^2 + \tilde{m}^2) (p'^2 + q^2 + \tilde{m}^2) \left(1 + \frac{\lambda}{\alpha} \sqrt{q^2 + \tilde{m}^2}\right)}.
\end{align}
According to Eq.~\eqref{eq:approximation-for-zz-1D}, we need the Green's function evaluated at $x = x' = 0$:
\begin{align}
\label{eq:greens-function-1D-continuum-fourier-transform}
\Galpha(0, y; 0, y') &= \int \frac{dp dq dp'}{(2 \pi)^3} e^{-i q (y - y')} \Galpha(p, q; p')\\
&= 2 \int_0^{\infty} d q \frac{\cos\left( q |y - y'|\right)}{\frac{\alpha}{\lambda} + \sqrt{q^2 + \tilde{m}^2}}.
\end{align}

By setting $\alpha = 0$ and expanding around $\tilde{m} = 0$, we obtain the
well known result
\begin{align}
\Galpha(0, y; 0, y')\big|_{\alpha = 0} &= 2 K_0(\tilde{m} |y - y'|) \sim -2 \ln\left(\tilde{m} |y - y'|\right)
\end{align}
for the correlator of the massless free boson.

For $\tilde{m} = 0$ and $\alpha > 0$, the Green's function becomes
\begin{align}
\Galpha(0, y; 0, y')\big|_{\tilde{m} = 0} &= -2 \left[\gaux{\frac{\alpha}{\lambda} |y - y'|}\right],
\end{align}
where $\Si$ and $\Ci$ are the sine and cosine integral functions, respectively.
Correspondingly, the approximation for the $zz$ correlator between the sites $l+1$ and $1$ with $l > 0$ becomes
\begin{align}
\zz{l+1}{1} &\approx 2 \alpha \left[\gaux{r}\right],
\end{align}
where $r = 2 \pi \alpha l$. Notice that $\zz{l+1}{1}$ is indeed independent
of the lattice scale $\lambda$.
Using the asymptotic expansion\cite{Gautschi1964}
\begin{align}
\gaux{r} \sim -\frac{1}{r^2},
\end{align}
we obtain the large-distance behavior
\begin{align}
\zz{l+1}{1} \sim -\frac{1}{2 \pi^2 \alpha} \frac{1}{l^2}.
\end{align}

Finally, let us consider a finite system of $N$ sites with
periodical boundary conditions. In this case,
the site index $n$ in Eq.~\eqref{eq:lattice-positions-1D-before-continuum-limit}
assumes values $n \in \{1, \dots, N\}$ and the continuum
limit of $\frac{\alpha}{2}\sum_{n=1}^N \phi(x_n, y_n)^2$ becomes
\begin{align}
\frac{\alpha}{4 \pi \lambda} \int_0^{L} dy \phi(0, y)^2,
\end{align}
where $L = 2 \pi \lambda N$ is the size of the system.
The computation of the Green's function in momentum space is
as for the case of the infinite system except that the
integral over $q$ is now replaced by a sum over discrete
momenta:
\begin{align}
\int \frac{dq}{2 \pi} \to \frac{1}{L} \sum_{q_k},
\end{align}
where $q_k = 2 \pi k/L$,
and $k \in \mathbb{Z}$. Therefore, the Green's function
evaluated at $x = x' = 0$ and $\tilde{m} = 0$ becomes
\begin{align}
G^{\alpha}(0, y; 0, y') &=
\frac{2\pi}{L} \sum_{k=-\infty}^{\infty} \frac{e^{-i q_k (y - y')}}{\frac{\alpha}{\lambda} + |q_k|} = - \frac{2 \pi \lambda}{L \alpha} + 2 \mathrm{Re}\left[\Phi\left(e^{\frac{2 \pi i}{L} (y - y')}, 1, \frac{L \alpha}{2 \pi \lambda}\right)\right],
\end{align}
where $\Phi(z, s, a)$ is the Lerch transcendent function defined
in Table~\ref{tab:summary-continuum-limit}. According
to Eq.~\eqref{eq:approximation-for-zz-1D},
we obtain
\begin{align}
\zz{l+1}{1} &\approx \frac{1}{N} - 2 \alpha \mathrm{Re}\left[\Phi\left(e^{2 \pi i \frac{l}{N}}, 1, N \alpha\right)\right].
\end{align}
Note that this result is again independent of the lattice scale.

\subsubsection{Two-dimensional system without boundary (sphere)}
\label{sec:appendix-condintuum-approximation-sphere}
On the sphere, the quadratic action that provides an approximation of
the $zz$ correlations is given by
\begin{align}
\Salpha &= \frac{1}{8 \pi} \int_{S^2} d \Omega \phi(\Omega) L^2 \phi(\Omega) +
\frac{\alpha}{2} \sum_{j=1}^N \phi(\Omega_j)^2,
\end{align}
where $L^2$ is the square of the orbital angular momentum operator (cf. appendix~\ref{sec:appendix-free-boson-sphere}).

We consider an approximately uniform distribution of positions $\Omega_j = (\theta_j, \varphi_j)$ on the sphere. As a consequence, each solid angle element $\Delta \Omega$
contains approximately the same number of points $\Omega_j$ and we approximate
\begin{align}
\frac{4 \pi}{N} \sum_{j=1}^N \phi(\Omega_j)^2 &\approx \int_{S^2} d \Omega \phi(\Omega)^2,\\
\Salpha &\approx \frac{1}{8 \pi} \int_{S^2} d \Omega \phi(\Omega) (L^2 + \alpha N) \phi(\Omega).
\end{align}
Using the result for the propagator of the massive boson on the sphere of Eq.~\eqref{eq:greens-function-sphere-massive},
the following approximation for the $zz$ correlations is obtained:
\begin{align}
\zz{i}{j} &\approx -\alpha \int_{0}^{\infty} d q \frac{2 \cos\left(\sqrt{N \alpha - \frac{1}{4}} q\right)}{\sqrt{2 \cosh(q) - 2 + |\spherepos{i} - \spherepos{j}|^2 }}.
\end{align}

\subsubsection{Two-dimensional system with a boundary}
Let us consider an infinite square lattice in the half-plane $x \ge 0$:
\begin{align}
x_{i_x} &=  \sqrt{2 \pi \lambda} (i_x - 1)\\
y_{i_y} &= \sqrt{2 \pi \lambda} i_y,
\end{align}
where $i_x \in \{1, 2, \dots\}$, $i_y \in \mathbb{Z}$, and $\sqrt{2 \pi \lambda}$ is the lattice constant ($\lambda > 0$).

The term in the action~\eqref{eq:quadratic-action-with-mass-regulator} that is proportional to $\alpha$ becomes
\begin{align}
\frac{\alpha}{2} \sum_{i_x=0}^{\infty} \sum_{i_y=-\infty}^{\infty} \phi(\sqrt{2 \pi \lambda} i_x, \sqrt{2 \pi \lambda} i_y)^2 \approx \frac{\alpha}{4 \pi \lambda} \int_{x \ge 0} dx dy \phi(x, y)^2,
\end{align}
where we have applied a continuum approximation to the sum.

The Green's function corresponding to the action
\begin{align}
\Salpha &= \frac{1}{8 \pi} \int dx dy \phi(x, y) \left[-\partial_x^2 - \partial_y^2 + \tilde{m}^2 + 2 \frac{\alpha}{\lambda} \theta(x) \right] \phi(x, y)
\end{align}
satisfies
\begin{align}
\left[-\partial_x^2 - \partial_y^2 + \tilde{m}^2 + 2 \frac{\alpha}{\lambda} \theta(x) \right] \Galpha(x, x'; y, y')&= 4 \pi \delta(x - x') \delta(y - y'),
\end{align}
where $\theta(x)$ is the Heaviside step function.

The Fourier transform $\Galpha(p, q; p')$ defined through Eq.~\eqref{eq:Greens-function-in-Fourier-space} satisfies
\begin{align}
(p^2 + q^2 + \tilde{m}^2) \Galpha(p, q; p') + \frac{\alpha}{\lambda} \int \frac{d k}{\pi i} \frac{\Galpha(k, q; p')}{k - p - i \epsilon} &= 8 \pi^2 \delta(p + p'),
\end{align}
where we have used the integral representation
\begin{align}
\theta(x) &= \int \frac{dk}{2 \pi i} \frac{e^{i k x}}{k - i \epsilon}\quad \left(\epsilon > 0 \text{ infinitesimal}\right).
\end{align}
Next, we make a change of variables from $\Galpha(p, q; p')$ to $F(p, q; p')$
as defined in Eq.~\eqref{eq:definition-of-F-from-G} and find
\begin{align}
\label{eq:F-in-2D}
F(p, q; p') - 2 \frac{\alpha}{\lambda} \frac{1}{i (p + p'+ i \epsilon)} + \frac{\alpha}{\lambda} \int \frac{d k}{\pi i} \frac{F(k, q; p')}{\left(k^2 + E_q^2 \right) \left(k - p - i \epsilon \right)} = 0,
\end{align}
where $E_q^2 = q^2 + \tilde{m}^2$.
This equation implies that $F(p, q; p')$ for fixed
$q$ does not have poles for $p$ and $p'$ in the upper half plane
[$\mathrm{Im}(p) \ge 0$ and $\mathrm{Im}(p') \ge 0$].
Under this condition, the integral in Eq.~\eqref{eq:F-in-2D}
can be closed in the upper half plane and computed
through the sum of residues of the integrand at $k = i E_q$ and $k = p + i \epsilon$.
This results in the equation
\begin{align}
\label{eq:F-in-2D-after-integration}
\frac{p^2 + \tilde{E}_q^2}{p^2 + E_q^2} F(p, q; p') = \frac{\alpha}{\lambda} \frac{F(i E_q, q, p')}{E_q (E_q + i p)} + 2 \frac{\alpha}{\lambda} \frac{1}{i (p + p' + i \epsilon)},
\end{align}
where $\tilde{E}_q^2 = E_q^2 + 2 \frac{\alpha}{\lambda}$.
The left hand side of this equation vanishes for $p = i \tilde{E}_q$ since
$F(p, q; p')$ does not have a pole at $p = i \tilde{E}_q$.
Setting additionally $p' = i E_q$ in Eq.~\eqref{eq:F-in-2D-after-integration},
we get
\begin{align}
\label{eq:sol-F-2D-1}
F(i E_q, q, i E_q) &= 2 \frac{E_q (E_q - \tilde{E}_q)}{E_q + \tilde{E}_q}.
\end{align}
Substituting this result in Eq.~\eqref{eq:F-in-2D-after-integration}
for $p'=i E_q$ allows us to solve for $F(p, q, i E_q)$,
\begin{align}
\label{eq:sol-F-2D-2}
F(p, q, i E_q) &= - 4 i \frac{\alpha}{\lambda} \frac{E_q}{(p + i \tilde{E}_q) (E_q + \tilde{E}_q)}.
\end{align}
With $F(i E_q, q; p') = F(p, q; i E_q)$, which follows from
$\Galpha(x, y; x', y') = \Galpha(x', y; x, y')$,
we use Eq.~\eqref{eq:sol-F-2D-2} in Eq.~\eqref{eq:F-in-2D-after-integration}
and obtain
\begin{align}
\label{eq:sol-F-2D-3}
F(p, q; p') &= -2 i \frac{\alpha}{\lambda}
\frac{(p + i E_q) (p' + i E_q)}
{(p + p' + i \epsilon) (p + i \tilde{E}_q) (p' + i \tilde{E}_q)}.
\end{align}
This result is indeed symmetric in $p$ and $p'$ and does not have poles in
the upper half plane.
Next, we need to integrate the Fourier-transformed Green's function
\begin{align}
\Galpha(p, q;, p') &= 8 \pi^2 \frac{\delta(p + p')}{p^2 + E_q^2} +
8 \pi \frac{\alpha}{\lambda} \frac{(p + i E_q) (p' + i E_q)}
{i (p + p' + i \epsilon) (p^2 + E_q^2) (p'^2 + E_q^2) (p + i \tilde{E}_q) (p' + i \tilde{E_q})}
\end{align}
back to coordinate space. We are interested in the
correlating along the $y$ direction for a fixed value of $x$.
This is why we choose $x' = x$:
\begin{align}
\Galpha(x, y; x, y') &= \int \frac{d p d q d p'}{(2 \pi)^2} e^{-i p x - i p' x - i q (y - y')} \Galpha(p, q; p')\\
&= 2 \int_0^{\infty} d q \frac{\cos\left(q \sqrt{2 \frac{\alpha}{\lambda}} |y - y'| \right)}{\sqrt{1 + q^2}} \left[1 + \frac{e^{-2 \sqrt{2 \frac{\alpha}{\lambda}} \sqrt{1 + q^2} x}}{\left(q + \sqrt{1 + q^2}\right)^2}\right].
\end{align}

An approximation of the correlator between sites
$(i_x, l+1)$ and $(i_x, 1)$ with $l > 0$ is then given by
\begin{align}
\zz{i_x, l + 1}{i_x, 1} &\approx -\alpha \Galpha\left(\sqrt{2 \pi \lambda} (i_x-1), \sqrt{2 \pi \lambda} l; \sqrt{2 \pi \lambda} (i_x-1), \sqrt{2 \pi \lambda}\right)\\
&= -2 \alpha \int_0^{\infty} d q \frac{\cos\left(2 \sqrt{\pi \alpha} q l \right)}{\sqrt{1 + q^2}} \left[1 + \frac{e^{-4 \sqrt{\pi \alpha} \sqrt{1 + q^2} (i_x - 1)}}{\left(q + \sqrt{1 + q^2}\right)^2}\right].
\end{align}
Notice that this result is again independent
of the lattice constant.

At the edge ($i_x = 1$), we get
\begin{align}
\zz{1, l+1}{1, 1} &\approx -4 \alpha \int_{0}^{\infty} dq \frac{\cos\left(2 \sqrt{\pi \alpha} q l\right)}{q + \sqrt{1 + q^2}}.
\end{align}
After two integrations by part, we obtain
\begin{align}
\zz{1, l+1}{1, 1} &\approx -\frac{1}{\pi l^2} + 2 \sqrt{\frac{\alpha}{\pi}} \frac{K_1(2 \sqrt{\pi \alpha} l)}{l},
\end{align}
where $K_n$ are the modified Bessel functions of the second kind.
For large distances $l \gg 1$, the second term
is exponentially suppressed so that
\begin{align}
\zz{1, l+1}{1, 1} \sim -\frac{1}{\pi l^2},
\end{align}
i.e. the correlations decay with a power of $-2$ independent of $\alpha$.

In the bulk ($i_x \to \infty$), we have
\begin{align}
\zz{i_x, l + 1}{i_x, 1}\big|_{i_x \to \infty} &\approx -2 \alpha \int_0^{\infty} d q \frac{\cos\left(2 \sqrt{\pi \alpha} q l \right)}{\sqrt{1 + q^2}}\\
&= -2 \alpha K_0(2 \sqrt{\pi \alpha} l).
\end{align}
For large distances $l \gg 1$
\begin{align}
\zz{i_x, l + 1}{i_x, 1}\big|_{i_x \to \infty} \sim -\pi^{\frac{1}{4}} \alpha^{\frac{3}{4}} \frac{e^{-2 \sqrt{\pi \alpha} l}}{\sqrt{l}},
\end{align}
i.e. the correlations decay exponentially with a correlation
length of $1/(2 \sqrt{\pi \alpha})$.

We note that one can put the system on a half-infinite cylinder instead
of the half-plane. Similar to the case of the 1D system,
this implies that the integral over $q$ is replaced by a sum over discrete momenta.
The resulting approximation for the $zz$ correlations is given by
\begin{align}
\label{eq:zz-approx-half-infinite-cylinder}
\zz{i_x, l + 1}{i_x, 1} &\approx
-\frac{2 \pi  \alpha}{N_y} \sum_{k=-\infty}^{\infty} e^{-i q_k l} \frac{1}{\sqrt{4 \pi  \alpha +q_k^2}}  \left[1 + \frac{4 \pi  \alpha  e^{-2 \left(i_x-1\right) \sqrt{4 \pi  \alpha +q_k^2}}}{\left(\left| q_k\right| +\sqrt{4 \pi  \alpha +q_k^2}\right)^2}\right],
\end{align}
where $q_k = 2 \pi k/N_y$ and $N_y$ is the number of sites in the periodical
direction of the cylinder. For the plots shown in the Supplementary Material\cite{supplement},
we evaluated Eq.~\eqref{eq:zz-approx-half-infinite-cylinder} numerically by setting
a cutoff momentum, doing a fast Fourier transform of the resulting finite sum, and
extrapolating to the case of an infinite cutoff momentum.

\subsection{Discrete approximation}
\label{sec:appendix-discrete-approximation}
\subsubsection{Discrete approximation from CFT operators}
Let us define the two-point Green's function of the discrete approximation as
\begin{align}
\label{eq:discrete-Greens-function-2pt}
\Galpha_{i, j} &= \frac{
\langle :\phi(y_i) e^{-\frac{\alpha}{2} \phi(y_i)^2}: :\phi(y_j) e^{-\frac{\alpha}{2} \phi(y_j)^2}: \prod_{k (\neq i, j)}^N :e^{-\frac{\alpha}{2} \phi(y_k)^2}: \rangle}
{\langle \prod_{k=1}^N :e^{-\frac{\alpha}{2} \phi(y_k)^2}: \rangle}
,
\end{align}
where the expectation value is taken with respect to action of the massless boson.
The generalized coordinate $y_j$ is assumed to be one of the following:
$y_j=(z_j, \bar{z}_j)$ for positions in the complex plane,
$y_j= \Omega_j = (\theta_j, \varphi_j)$
on the sphere, and $y_j=(w_j, \bar{w}_j)$ on the cylinder.
The $zz$ correlation in the discrete approximation is then given by
\begin{align}
\zz{i}{j} &\approx -\alpha \Galpha_{i, j}.
\end{align}

The following computation makes use of the correlator of $N$ vertex operators
given by
\begin{align}
\label{eq:correlator-of-vertex-operators-general-coordinate}
\langle \prod_{j=1}^N :e^{i \sqrt{\alpha} s_j \phi(y_j)}: \rangle &= \delta_{\mathbf{s}} \prod_{m < n}^N d_{m, n}^{2 \alpha s_m s_n},
\end{align}
where
\begin{align}
\label{eq:d-definition-for-lattices}
d_{m, n} &= \begin{cases}
|z_m - z_n|& \text{(complex plane)},\\
|\spherepos{m} - \spherepos{n}|& \text{(sphere)},\\
|2 \sinh(\frac{1}{2} (w_m - w_n))|& \text{(cylinder)},
\end{cases}
\end{align}
cf. Appendix~\ref{sec:appendix-free-boson-sphere-cylidner} for the computation
on the sphere and on the cylinder. For the calculation below, it is important to
note that
Eq.~\eqref{eq:correlator-of-vertex-operators-general-coordinate} holds
not only for $s_j \in \{-1, 1\}$ but for general $s_j \in \mathbb{R}$.

We define higher order Green's functions $\Galpha_{i_1, \dots, i_{2n}}$ analogously
to Eq.~\eqref{eq:discrete-Greens-function-2pt}. Let us denote the generating
function of $\Galpha_{i_1, \dots, i_{2n}}$
by $\Zalpha(\mathbf{J})$, where $\mathbf{J}$ is an $N$-dimensional,
real vector:
\begin{align}
  \Galpha_{i_1, \dots, i_{2n}} &=\left(-\frac{1}{\alpha}\right)^{n} \frac{1}{\Zalpha(\mathbf{J})} \frac{\partial}{\partial J_1} \frac{\partial}{\partial J_2} \dots \frac{\partial}{\partial J_{2n}} \Zalpha(\mathbf{J})\Big|_{\mathbf{J} = 0},\\
  \label{eq:definition-of-generating-Zalpha}
  \Zalpha(\mathbf{J}) &= \langle \prod_{j=1}^N :e^{-\frac{\alpha}{2} \phi(y_j)^2 + i \sqrt{\alpha} J_j \phi(y_j)}: \rangle.
\end{align}

The generating function $\Zalpha(\mathbf{J})$ will now be evaluated using
the $N$-dimensional Gaussian integral~\cite{DiFrancesco1997}
\begin{align}
  \label{eq:N-dimensional-Gaussian-integral}
  \int d^N \mathbf{x} e^{-\frac{1}{2} \mathbf{x}^t A \mathbf{x} + \mathbf{b}^t \mathbf{x}} &= \left(\frac{(2 \pi)^N}{\det A}\right)^{\frac{1}{2}} e^{\frac{1}{2} \mathbf{b}^t A^{-1} \mathbf{b}},
\end{align}
where $A$ is a symmetric, real $N \times N$ matrix with
positive eigenvalues and $\mathbf{b}$ a real vector of length $N$.
By Fourier transforming
\begin{align}
  :e^{-\frac{1}{2} \alpha \phi(y_j)^2}: &= \frac{1}{\sqrt{2 \pi}} \int_{-\infty}^{\infty} d s_j e^{-\frac{1}{2} s_j^2} :e^{i \sqrt{\alpha} s_j \phi(y_j)}:,
\end{align}
we obtain
\begin{align}
  \Zalpha(\mathbf{J}) 
 &\propto \int d^N \mathbf{s} e^{-\frac{1}{2} \mathbf{s}^2} \langle \prod_{j=1}^N :e^{i \sqrt{\alpha} (s_j + J_j) \phi(y_j)}: \rangle.
\end{align}
Using Eq.~\eqref{eq:correlator-of-vertex-operators-general-coordinate} results in
\begin{align}
  \Zalpha(\mathbf{J}) &\propto \int d^N \mathbf{s} e^{-\frac{1}{2} \mathbf{s}^2} \delta\left(\sum_{j=1}^N s_j + \sum_{j=1}^N J_j\right) \prod_{m < n}^N d_{m, n}^{2 \alpha (s_m + J_m) (s_n + J_n)}\\
                            &= \int d^N \mathbf{s} e^{-\frac{1}{2} \left(\mathbf{s} - \mathbf{J}\right)^2} \delta\left(\sum_{j=1}^N s_j\right) \prod_{m < n} d_{m, n}^{2 \alpha s_m s_n}\\
\label{eq:Zalpha-as-integral}
&= e^{-\frac{1}{2} \mathbf{J}^2} \int d^N \mathbf{s}
\delta\left(\sum_{j=1}^N s_j\right) e^{-\frac{1}{2} \mathbf{s}^t M
\mathbf{s} + \mathbf{J}^t \mathbf{s}},
\end{align}
where
\begin{align}
\label{eq:definition-matrix-M}
M_{m, n} &= \Mmat{m}{n}.
\end{align}
One of the $N$ integrals can be evaluated due to the $\delta$
function. To this end, we choose some index $r \in \{1, \dots, N\}$
and introduce the $N \times N$ matrix $T_r$ by the linear
transformation
\begin{align}
  T_r \mathbf{s} &= \left(\begin{matrix}
      s_1\\
      \vdots\\
      s_{r-1}\\
      -\sum_{j(\neq r)} s_j\\
      s_{r+1}\\
      \vdots\\
      s_N
      \end{matrix}
  \right),
\end{align}
i.e. the matrix entries of $T_r$ are given by
\begin{align}
  \left(T_r\right)_{m, n} &= \Trmat{m}{n}.
\end{align}
Carrying out the integral over $s_r$ results in
\begin{align}
  \Zalpha(\mathbf{J})
  \propto e^{-\frac{1}{2} \mathbf{J}^2}
  \int \left(\prod_{j(\neq r)} d s_j\right)
  e^{-\frac{1}{2} \mathbf{s}^t T_r^t M T_r \mathbf{s} + \mathbf{J}^t T_r \mathbf{s}}.
\end{align}
Using
\begin{align}
  \sqrt{2 \pi} =
  \int_{-\infty}^{\infty} d s_r e^{-\frac{1}{2} s_r^2} = \int_{-\infty}^{\infty} d s_r e^{-\frac{1}{2} \mathbf{s}^t \evec{r}\evec{r}^t \mathbf{s}},
\end{align}
where $\evec{r}$ denotes the $r$th unit vector,
we reintroduce an integral over $s_r$ and obtain
\begin{align}
  \Zalpha(\mathbf{J})
  &\propto e^{-\frac{1}{2} \mathbf{J}^2} \int d^N \mathbf{s}
  e^{-\frac{1}{2} \mathbf{s}^t (T_r^t M T_r + \evec{r} \evec{r}^t)
  \mathbf{s} + \mathbf{J}^t T_r \mathbf{s}}.
\end{align}
With the $N$-dimensional Gaussian integral of Eq.~\eqref{eq:N-dimensional-Gaussian-integral}, we
arrive at
\begin{align}
  \Zalpha(\mathbf{J}) \propto e^{\frac{1}{2} \mathbf{J}^t \Gamma \mathbf{J}},
\end{align}
where
\begin{align}
  \Gamma = \Gammamat.
\end{align}
Note that $\Gamma$ is independent of the choice of the index $r$
since it does not matter which of the integrals is evaluated using
the $\delta$ function.

It follows that
\begin{align}
  \Galpha_{i, j} &=
  -\frac{1}{\alpha} \frac{\partial}{\partial J_i} \frac{\partial}{\partial J_j} e^{\frac{1}{2} \mathbf{J}^t \Gamma \mathbf{J}}\big|_{\mathbf{J} = 0}
  = -\frac{1}{\alpha} \Gamma_{i,j}
  \intertext{and}
  \zz{i}{j} &\approx \Gamma_{i,j}.
\end{align}

Let us now compute the subleading term in the expansion of $\zz{i}{j}$.
To this end, we write the exact $zz$
correlations as
\begin{align}
\zz{i}{j} &=
- \frac{
\langle
\prod_{k \in \{i, j\}} :\tan(\sqrt{\alpha} \phi(y_k)) C(\sqrt{\alpha} \phi(y_k)) e^{-\frac{\alpha}{2} \phi(y_k)^2}:
\prod_{k(\neq i, j)}^N :C(\sqrt{\alpha} \phi(y_k)) e^{-\frac{\alpha}{2} \phi(y_k)^2}:
\rangle
}
{\langle \prod_{k=1}^N :C(\sqrt{\alpha} \phi(y_k)) e^{-\frac{\alpha}{2} \phi(y_k)^2}: \rangle},
\end{align}
where $C(x)
= \cos(x)  e^{\frac{1}{2} x^2}$. In terms of the generating function $\Zalpha$
of Eq.~\eqref{eq:definition-of-generating-Zalpha},
\begin{align}
\label{eq:exact-zz-from-generating-Zalpha}
\zz{i}{j} &= -\frac{
\tan(-i \partial_i) \tan(-i \partial_j)
\prod_{k=1}^N C(-i \partial_k) \Zalpha(\mathbf{J})
\big|_{\mathbf{J} = 0}}
{
\prod_{k=1}^N C(-i \partial_k) \Zalpha(\mathbf{J})\big|_{\mathbf{J} = 0}},
\end{align}
where $\partial_k = \frac{\partial}{\partial{J_k}}$.
Expanding to fourth order in derivatives,
\begin{align}
\tan(-i \partial_i) \tan(-i \partial_j)
\prod_{k=1}^N C(-i \partial_k) &=
-\partial_i \partial_j +
\frac{1}{3} (\partial_i^3 \partial_j + \partial_i \partial_j^3) + \dots,\\
\prod_{k=1}^N C(-i \partial_k) &= 1 - \frac{1}{12} \sum_{k=1}^N \partial_k^4 + \dots,
\end{align}
we obtain
\begin{align}
\zz{i}{j} &= \Gamma_{i, j} - \Gamma_{i, j} (\Gamma_{i, i} + \Gamma_{j, j}) + \dots.
\end{align}

\subsubsection{Alternative derivation of discrete approximation}
We now demonstrate that the discrete approximation can be obtained
directly from the expression of the exact $zz$ correlations in
$\psi_{\alpha}$ without using CFT operators.

Let us first consider the norm squared of $\psi_{\alpha}$:
\begin{align}
\langle \psi_{\alpha} | \psi_{\alpha} \rangle &= \sum_{s_1, \dots, s_N} \delta_{\mathbf{s}} \prod_{m < n} d_{m, n}^{2 \alpha s_m, s_n} =
\sum_{s_1, \dots, s_N} \delta_{\mathbf{s}} e^{-\frac{1}{2} \mathbf{s}^t (M - \mathbb{I}) \mathbf{s}},
\end{align}
where $d_{m, n}$ is defined in Eq.~\eqref{eq:d-definition-for-lattices}
and $M$ in Eq.~\eqref{eq:definition-matrix-M}.

Writing the sums over each $s_j$ as an integral,
\begin{align}
\sum_{s_j \in \{-1, 1\}} \to \int_{-\infty}^\infty d s_j \left(\delta(s_j + 1) + \delta(s_j - 1)\right),
\end{align}
we obtain
\begin{align}
\langle \psi_{\alpha} | \psi_{\alpha} \rangle &\propto
\int d^N \mathbf{s} \left[\prod_{m=1}^N  \left(\delta(s_m + 1) + \delta(s_m - 1)\right)\right] \delta\left(\sum_{m=1}^N s_m\right) e^{-\frac{1}{2} \mathbf{s}^t (M - \mathbb{I}) \mathbf{s}}\\
&\propto \int d^N \mathbf{x} \left[\prod_{m=1}^N \cos(x_m)\right] \int d^N \mathbf{s} e^{-i \mathbf{x}^t \mathbf{s}} \delta\left(\sum_{m=1}^N s_m\right) e^{-\frac{1}{2} \mathbf{s}^t (M - \mathbb{I}) \mathbf{s}},
\end{align}
where it was used that
\begin{align}
\delta(s_m + 1) + \delta(s_m - 1) &= \int_{-\infty}^{\infty} \frac{d x_m}{\pi} \cos(x_m) e^{-i x_m s_m}.
\end{align}
For the expectation value
$4 \langle \psi_{\alpha} | t^z_i t^z_j | \psi_{\alpha} \rangle$,
we additionally replace
\begin{align}
\sum_{s_j \in \{-1, 1\}} s_j &\to \int_{-\infty}^\infty d s_j \left(\delta(s_j - 1) - \delta(s_j + 1)\right)
\intertext{and use}
\delta(s_j - 1) - \delta(s_j + 1) &=  i \int_{-\infty}^{\infty} \frac{d x_j}{\pi} \sin(x_j) e^{-i x_j s_j}
\end{align}
so that
\begin{align}
4 \langle \psi_{\alpha} | t^z_i t^z_j | \psi_{\alpha} \rangle &\propto -\int d^N \mathbf{x} \tan(x_i) \tan(x_j) \left[\prod_{m=1}^N \cos(x_m)\right] \int d^N \mathbf{s} e^{-i \mathbf{x}^t \mathbf{s}} \delta\left(\sum_{m=1}^N s_m\right) e^{-\frac{1}{2} \mathbf{s}^t (M - \mathbb{I}) \mathbf{s}}.
\end{align}

Let us introduce the generating function\begin{align}
\label{eq:definition-Zalphap}
\Zalpha'(\mathbf{J}) &= \int d^N \mathbf{x} e^{-\frac{1}{2} \mathbf{x}^2 + i \mathbf{J}^t \mathbf{x}} 
\int d^N \mathbf{s} e^{-i \mathbf{x}^t \mathbf{s}} \delta\left(\sum_{j=1}^N s_j\right) e^{-\frac{1}{2} \mathbf{s}^t (M - \mathbb{I}) \mathbf{s}},
\end{align}
which is defined in such a way that the exact $zz$ correlations $\zz{i}{j} = 4 \frac{\langle \psi_{\alpha} | t^z_i t^z_j | \psi_{\alpha} \rangle}{\langle \psi_{\alpha} | \psi_{\alpha} \rangle}$ are given by Eq.~\eqref{eq:exact-zz-from-generating-Zalpha} with $\Zalpha(\mathbf{J})$
replaced by $\Zalpha'(\mathbf{J})$.
In particular, the $zz$ correlations in the quadratic approximation are given
by
\begin{align}
\zz{i}{j} \approx \frac{1}{\Zalpha'(\mathbf{J})} \frac{\partial}{\partial J_i}  \frac{\partial}{\partial J_j} \Zalpha'(\mathbf{J})\big|_{\mathbf{J} = 0}.
\end{align}

We now show that $\Zalpha'(\mathbf{J})$ is equivalent to the generating function
$\Zalpha(\mathbf{J})$ of the previous derivation.
Carrying out the integral over $\mathbf{x}$ in
Eq.~\eqref{eq:definition-Zalphap},
\begin{align}
\Zalpha'(\mathbf{J}) &\propto \int d^N \mathbf{s} e^{-\frac{1}{2} (\mathbf{J} - \mathbf{s})^2} \delta\left(\sum_{j=1}^N s_j\right) e^{-\frac{1}{2} \mathbf{s}^t (M - \mathbb{I}) \mathbf{s}}\\
&= e^{-\frac{1}{2} \mathbf{J}^2}\int d^N \mathbf{s} \delta\left(\sum_{j=1}^N s_j\right) e^{-\frac{1}{2} \mathbf{s}^t M \mathbf{s} + \mathbf{J}^t \mathbf{s}}.
\end{align}
Comparing to $\Zalpha(\mathbf{J})$ in Eq.~\eqref{eq:Zalpha-as-integral}, we conclude
that $\Zalpha'(\mathbf{J}) \propto \Zalpha(\mathbf{J}) \propto e^{\frac{1}{2} \mathbf{J}^t \Gamma \mathbf{J}}$.
\end{widetext}

\bibliography{refs}

\end{document}